\documentclass[aps,prb,twocolumn,longbibliography,superscriptaddress]{revtex4-1}

\usepackage[dvips]{graphicx}
\usepackage{amsmath,amssymb}
\usepackage{amsfonts}
\usepackage[table,usenames,dvipsnames]{xcolor}
\usepackage[retainorgcmds]{IEEEtrantools}
\usepackage{bbold}
\usepackage[normalem]{ulem}
\newcommand{\mycolor}{\color{black}} 

\begin{document}
\newcommand{\ui}{\text{i}}
\newcommand{\sech}{\;\text{sech}}
\newcommand{\Eq}[1]{Eq.~\eqref{#1}}
\newcommand{\Eqs}[1]{Eqs.~\eqref{#1}}

\newcommand{\IJbb}{I_c^{B}}
\newcommand{\CJbb}{C_J^{B}}
\newcommand{\IJab}{\hat I_c}
\newcommand{\CJab}{\hat C_J}
\newcommand{\IJq}{I_c^{q}}
\newcommand{\EJq}{E_J^{q}}
\newcommand{\LJq}{L_J^{q}}
\newcommand{\CJq}{C_J^{q}}
\newcommand{\Lq}{L_{q}}
\newcommand{\Ufl}{U_{\text{fl}}}
\newcommand{\Efl}{E_{\text{fl}}}
\newcommand{\Uq}{U_{\text{q}}}
\newcommand{\Vint}{V_{\text{int}}}
\newcommand{\Hq}{H_{q}}

\newcommand{\phiBB}{\phi^B}
\newcommand{\phiL}{\phi_L}
\newcommand{\phiR}{\phi_R}
\newcommand{\phiq}{\phi_{q}}
\newcommand{\phiqs}{\phi_{qs}}
\newcommand{\phiext}{\varphi_{\text{ext}}}
\newcommand{\pq}{\pi_{q}}
\newcommand{\nq}{n_{q}}

\newcommand{\CJeff}{C_{J,\text{eff}}}
\newcommand{\IJeff}{I_{c,\text{eff}}}
\newcommand{\Leff}{L_{\text{eff}}}
\newcommand{\phiLR}{\phi_{LR}}
\newcommand{\Mq}{M_{q}}
\newcommand{\Mb}{M_{B}}
\newcommand{\Mlr}{M_{LR}}

\newcommand{\Csh}{C_{\text{sh}}}

\newcommand{\flqu}{\frac{\Phi_0}{2\pi}}
\newcommand{\iflqu}{\frac{2\pi}{\Phi_0}}
\newcommand{\uGHz}{\,\text{GHz}}
\newcommand{\uMHz}{\,\text{MHz}}
\newcommand{\ums}{\,\text{ms}}
\newcommand{\uns}{\,\text{ns}}
\newcommand{\ups}{\,\text{ps}}
\newcommand{\umicroA}{\,\mu \text{A}}
\newcommand{\umicrom}{\,\mu \text{m}}
\newcommand{\ufF}{\,\text{fF}}
\newcommand{\unH}{\,\text{nH}}

\title{Simulation of a rapid qubit readout dependent on the transmission of a single fluxon}

\author{W.~Wustmann}
\affiliation{The Department of Physics, University of Otago, Dunedin 9016, New Zealand}
\affiliation{The Dodd-Walls Centre for Photonic and Quantum Technologies, University of Otago, Dunedin 9016, New Zealand}
\affiliation{The Laboratory for Physical Sciences at the University of Maryland,
College Park, MD 20740, USA}

\author{K.D.~Osborn}
\email{kosborn@umd.edu (corresponding author)}
\affiliation{The Laboratory for Physical Sciences at the University of Maryland,
College Park, MD 20740, USA}
\affiliation{The Quantum Materials Center, University of Maryland,
College Park, MD 20742, USA}
\affiliation{The Joint Quantum Institute, University of Maryland,
College Park, MD 20742, USA}

\date{\today}

\begin{abstract}
The readout speed of qubits is a major limitation for error correction in quantum information science.
We show simulations of a proposed device that gives readout of a fluxonium qubit using a ballistic fluxon with an estimated readout time of less than 1 nanosecond, without the need for an input microwave tone. This contrasts the prevalent readout based on circuit quantum electrodynamics, but is related to previous studies where a fluxon moving in a single long Josephson junction (LJJ) can exhibit a time delay depending on the state of a coupled qubit. Our readout circuit contains two LJJs and a qubit coupled at their interface.
We find that the device can exhibit single-shot readout of a qubit --- one qubit state leads to a single dynamical bounce at the interface and fluxon reflection, and the other qubit state leads to a couple of bounces at the interface and fluxon transmission.
Dynamics are initially computed with a separate degree of freedom for all Josephson junctions of the circuit.
However, a collective coordinate model reduces the dynamics to three degrees of freedom: one for the fluxonium Josephson junction and one for each LJJ.
The large mass imbalance in this model allows us to simulate the
mixed quantum-classical dynamics, as an approximation for the full quantum dynamics. Calculations give backaction on the qubit at $\leq 0.1\%$.
\end{abstract}


\maketitle 

\section{Introduction}

Qubit readout is often viewed as a performance-limiting process in quantum information science (QIS). In one of the first QIS platforms, superconducting qubits were read out using strong-coupled SQUIDs or single-electron transistors (SETs)\cite{Switch2, Switch3}.
However, in recent years, cavity quantum electrodynamics (c-QED) readouts \cite{Cavity1, Cavity2, cQED} have taken over as the prevalent type. In quantum information systems, readouts of several qubits are currently used in quantum error correction studies \cite{SchoelkopfEC, WallraffEC, GoogleEC}, and readout is currently one of the top two limitations in the error budget \cite{GoogleEC}. The readout time is typically 100s of nanoseconds, and an improved readout should have a higher speed and smaller backaction \cite{ChalmersReadout, msMemory}. 

Readout that uses a ballistic flux quantum in long Josephson junctions (LJJs) is less explored than the prevalent type. As initially proposed by Averin, Rabenstein and Semenov \cite{AveRabSem2006}, two fundamental readout modes
arise when a ballistic fluxon interacts locally with a qubit.
In a {\em time-delay mode} readout, the fluxon undergoes a time delay dependent on the qubit state during its transmission through the interaction region, which is equivalent to a weak measurement.
This readout method with fluxons has been explored in multiple theoretical studies \cite{FedorovETAL2007, AnnaHerr2007b, Ustinov2013, Kuzmin2015}, and
an experimental demonstration with an annular LJJ  \cite{Ustinov2014}, where the qubit state causes a frequency shift on the circulating fluxon.
On the other hand, a {\em transmission mode} readout is one where a fluxon transmits or reflects at the interaction region probabilistically, depending on the qubit state. This is analogous to a single-electron transistor reading a charge qubit \cite{SETreadout}, and this readout mode constitutes a strong measurement because it will collapse the qubit wave function.

In this study, we explore the {\em transmission-mode} readout of a qubit strongly coupled to two LJJs, where the result appears to be a usable fast single-shot readout. 
Simulations of the proposed device give a high-contrast readout {\mycolor with low backaction}, such that the transmission mode with two LJJs appears more achievable than previous work \cite{AveRabSem2006} with a single LJJ, which did not include simulations. 
For specific parameters, we observe that the fluxon-scattering outcome is sensitive to the qubit state --- one qubit state leads to reflection and the other to transmission.
The transmission case presented involves bounces of the fluxon at the interface between the two LJJs, where it interacts with the qubit differently than previous related work. During the scattering, the fluxon bounces up to a couple of times, where the time scale of the bounces is comparable to the plasma period in the LJJ. Our readout time, estimated below 1 nanosecond, is fast because of the ballistic nature of the dynamics, the high initial fluxon velocity, and the short scattering duration at the interface, even for a couple of bounces. 

When implemented, our device must be coupled to a fluxon launcher and detector, but our current study focuses on the interaction between the fluxon and a qubit. Two other readouts can also couple to superconducting electronics. One is the Josephson Photomultiplier \cite{JPM}, which has been demonstrated with a readout time of $\approx 100 \uns$, and the second is a developing readout based on the Quantum Flux Parametron \cite{MukhanovReadout}. Both readouts use a microwave input to generate cavity pointer states; in contrast, our qubit readout does not require a microwave input tone. 

This article is organized as follows.
{\mycolor
In Sec.~\ref{sec:circuit}, we present our circuit which strongly couples two independent LJJs and a fluxonium qubit to an interface cell. 
The interface cell allows complex bounce dynamics in the transmission mode because of the high quality of the barrier material. 
The section also summarizes properties of fluxons in LJJs, and discusses the parameter regimes suitable for the readout. 
Section \ref{sec:classical} presents two types of classical simulations -- a full circuit simulation and a collective coordinate model which allows us to reduce the many circuit degrees of freedom to only three. 
After describing how the initial condition in these simulations is set according to the two qubit states, we present the simulation results. 
In Sec.~\ref{sec:quantum} we then estimate the quantum backaction on the qubit, based on the quantum dynamics of the driven qubit. 
We also simulate the quantum dynamics of the coupled qubit-LJJ system based on the collective coordinate model. 
To this end, we use a mixed quantum-classical approach based on the large mass imbalance between the qubit and the two LJJ-degrees of freedom.
}

\section{Circuit}\label{sec:circuit}

\begin{figure*}[bt]
\includegraphics[width=\textwidth]{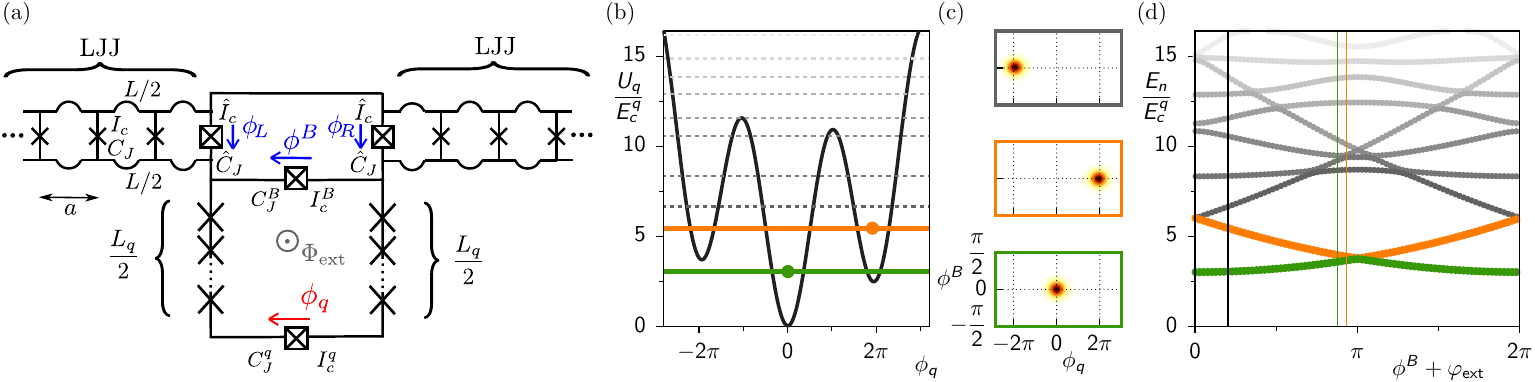}
\caption{
 (a)  Schematic of the device.
{\mycolor
Between two LJJs is the circuit interface cell consisting of three JJs connected in a loop: two `termination JJs’ (with phases $\phiL, \phiR$) and a `rail’-JJ (phase $\phiBB$). 
A fluxonium qubit is connected in parallel with the rail-JJ.
}
 It is composed of a small JJ (with phase $\phiq$)
 in series with a large `superinductance' from a large-JJ array.
 (b) Fluxonium potential $U_q(\phi_q)$ and first energy levels of the isolated
 fluxonium Hamiltonian $H_q$, \Eq{eq:Hq_phiq},
 for external flux $\phiext=0.2\pi$, and parameters $\IJq/I_c \approx 0.98$,
 $\CJq/C_J \approx 0.74$, $\Lq/L \approx 233$, and $\beta^2 = (4e^2/\hbar) \sqrt{L/C_J} = 0.4$,
 giving rise to the qubit transition energy of $\hbar \omega_{01} = 2.4 E_c^q = 0.17 E_0$.
 The characteristic fluxonium energies fulfill $E_J^q > 4 E_c^q \gg E_L^q$.
 (c) The first three eigenstates of
 the Hamiltonian for the two coupled phases $(\phiq,\phiBB)$, \Eq{eq:H2_boundstate},
 are strongly localized at $\phiBB \approx 0$,
 and in three distinct wells of $U_q(\phi_q)$.
 (d) Spectrum of the isolated fluxonium Hamiltonian $H_q$ under $\phiBB$-variation.
 Starting from $\phiBB \approx 0$ (solid vertical line), during the fluxon scattering, $\phiBB(t)$ temporarily reaches a maximum amplitude
 $\lesssim \pi$ (dashed vertical lines),
 e.g. $\text{max}(\phiBB) \approx 0.73\pi$ if the fluxonium is initialized in state $n=1$. The qubit is thus driven close to the avoided level crossing
 of states $n=0,1$ at $\phiBB + \phiext = \pi$, but nevertheless unintended
 state transfer remains very small, cf.~panels (iii) in Fig.~\ref{fig2}
 and discussion in Sec.~\ref{sec:backaction}.
}
\label{fig1}
\end{figure*}

\subsection{Fluxonium qubit at the interface of two LJJs}

In this study we have chosen a fluxonium qubit \cite{Manucharyan2009, Manucharyan2019},
but in principle a capacitively-shunted flux qubit could be used as well \cite{Mooij1999, Orlando1999, Orlando2016}.
Fluxonium qubits have improved considerably over the last years
and reach coherence times of up to one millisecond \cite{Manucharyan2019, ZhangETAl2021, ManucharyanETAL2021}.
A fluxonium consists of a small Josephson junction (JJ), with Josephson energy
$E_J^q = \IJq \Phi_0/(2\pi)$,
a parallel capacitor with charging energy $E_c^q=e^2/(2 \CJq)$,
and a parallel `superinductance' $L_q$,
which is produced by an array of Josephson junctions and contributes energy $E_L^q = (\hbar/2e)^2 /\Lq$.
An external flux bias $\Phi_{\text{ext}}=\phiext \cdot \Phi_0/(2\pi)$ through the qubit loop
can be used to tilt and manipulate the fluxonium potential,
where $\Phi_0 = h/(2e)$ is the flux quantum.

Isolated, the fluxonium is characterized by the Hamiltonian
\begin{align}
\label{eq:Hq_phiq}
H_q(\phiq) &= 4 E_c^q \nq^2 + U_q(\phiq) \\
\label{eq:Uq_phiq}
U_q(\phiq) &= E_J^q ( 1 - \cos\phiq ) + \frac{E_L^q}{2} (\phiq - \phiext)^2
\end{align}
where $\phiq$ is the phase over the small JJ,
and where the Cooper-pair number operator
 $\nq = \hbar/(2e)^2 \CJq \dot{\phi}_{q}$
satisfies $\left[ \phiq, \nq \right] = \ui$.
Typically, the fluxonium potential is presented in the alternative form
$U_q = E_J^q \left( 1 - \cos(\phi'_q + \phiext) \right) + E_L^q (\phi'_q)^2/2$,
which can be obtained through the gauge transformation
$\phi_q \rightarrow \phi'_q$
where $\phi'_q = \phi_q - \phiext$ is phase difference over the inductor
\cite{Manucharyan2019, Manucharyan2024}.

We start with the classical limit of the readout,
and for this reason, we require that the qubit flux states are macroscopically distinguishable.
This implies that the fluxonium should be designed and $\Phi_{\text{ext}}$-biased
such that $U_q$ has at least two wells separated
by approximately $2\pi$ in $\phiq$,
and that the ground state and first excited state lie in different potential wells, cf.~Fig.~\ref{fig1}(b,c).
Fluxonium circuits in the so-called `heavy fluxonium regime', $E_c^q, E_L^q \ll E_J^q$,
fulfill this criterium \cite{SchusterETAL2021} and we will discuss a readout
case for such a heavy fluxonium in App.~\ref{app:alternativereadout}.
The fluxonium, which we present in the body of this work,
is in the standard fluxonium regime, $E_L^q \ll 4 E_c^q > E_J^q$,
with flux bias $\Phi_{\text{ext}}$ close to an integer multiple \cite{Manucharyan2024} of $\Phi_0$.

Fluxons, here intended to probe the flux state of the qubit, can
be generated in underdamped long Josephson junctions (LJJs)
and move ballistically within them.
LJJs are JJs in which the dimension along the barrier is `long', i.e. exceeding
the Josephson penetration depth $\lambda_J$, the characteristic length
scale of flux excitations.
Instead of an ideal continuous LJJ, we consider here a discrete version \cite{HanLJJ1},
see Fig.~\ref{fig1}(a),
composed of JJs (with critical current $I_c$ and capacitance $C_J$)
interconnected by inductor `rails' with total inductance $L$ per cell, which has length $a$.
The array discreteness is characterized by the ratio $a/\lambda_J = \sqrt{L/L_J}$.
{\mycolor
We chose to use discrete LJJs rather than continuous ones, because in principle it allows one to fabricate the circuit using component inductors and Josephson junctions for a flexible design. 
Previous studies of fluxon dynamics in LJJs have been based on discrete LJJs for the same reason \cite{HanLJJ1}.
}

In our readout circuit,
the coupling of the fluxonium to the LJJs, and of the LJJs to each other,
involves an interface cell, see Fig.~\ref{fig1}(a).
This circuit is related to one of our previously studied classical ballistic gates,
which acts like a ballistic flip-flop (BFF) gate \cite{WusOsb2023_BSR},
the functional unit of a ballistic shift register.
The BFF contains an interface cell between two LJJs and a flux storage loop connected to the interface cell.
Our readout circuit contains the fluxonium in place of the flux storage loop.
By connecting the qubit galvanically to the LJJ interface cell,
the qubit phase and LJJ degrees of freedom are strongly coupled.
This differs from the coupling of previous works \cite{AveRabSem2006, AnnaHerr2007b, FedorovETAL2007, Kuzmin2015}
where the qubit is locally coupled to an LJJ, e.g. to a single cell of a discrete LJJ,
and the resulting mutual or shared inductance is small.
Due to the small interaction strength in those works, readout would typically be a weak measurement, using a sequence of many fluxons.

\subsection{The Langrangian}

The full system, shown in Fig.~\ref{fig1}(a), consists of left and right LJJ,
and the interface with a strongly coupled qubit.
The classical equations of motion can be obtained from its Lagrangian
\begin{align}
\label{eq:Lagrangian}
 \mathcal{L} &= \mathcal{L}_I + \mathcal{L}_l + \mathcal{L}_r \\
\label{eq:Lagr_intf}
\mathcal{L}_I
 &= \frac{\hbar^2}{8e^2} \left[
 \CJab  \left(\dot{\phi}_L^2 + \dot{\phi}_R^2 \right)
 + \CJbb  \left(\dot{\phi}^B\right)^2 \right]
  \\
 &
 + \frac{\IJab \Phi_0}{2\pi} \left( \cos\phiL + \cos\phiR \right)
 + \frac{\IJbb \Phi_0}{2\pi} \cos\phiBB
 \nonumber \\
 &
 + \frac{\hbar^2\CJq}{8 e^2}  (\dot{\phi}_{q})^2
 -\frac{ E_L^q}{2} \left(\phiq - \phiext - \phiBB\right)^2
 + \EJq \cos\phiq  \nonumber
\end{align}
Here, $\phiL := \phi^{(l)}_{k=0}$ and $\phiR := \phi^{(r)}_{k=0}$
are shorthand notations for the phases of the left and right interface JJs, respectively.
Since they terminate the respective LJJ (at nodes $k=0$, cf.~\Eq{eq:Lagr_LJJ}),
we call these the `termination JJs' of the interface, whereas the interface JJ connecting the lower rails of the two LJJs
is called the `rail JJ'.
The phases $\phiL, \phiR, \phiBB$ of these interface JJs
are constrained by flux quantization in the interface cell,
 $\phiBB = \phiL - \phiR$.
In \Eq{eq:Lagr_intf}, the coupling energy between the
fluxonium and the interface is given by
\begin{align}
\label{eq:Vint}
\Vint &= -E_L^q \phiBB (\phiq - \phiext)
 \,.
\end{align}
We design the LJJs as discrete structures,
consisting of individual JJs with critical current $I_c$ and capacitance $C_J$,
and connected by inductor rails with total cell inductance $L$.
The contributions of the two LJJs ($i=l,r$)
to the Langrangian, \Eq{eq:Lagrangian}, then read
\begin{align}\label{eq:Lagr_LJJ}
\mathcal{L}_i
&= \sum_{k \geq 1} \biggl[
\frac{\hbar^2 C_J}{8e^2}  \left(\dot{\phi}_k^{(i)}\right)^2 \Bigr.\\
& \Bigl.
- \frac{\Phi_0^2}{8 \pi^2 L} \left(\phi_{k}^{(i)} - \phi_{k-1}^{(i)}\right)^2
- \frac{I_c\Phi_0}{2\pi} \left(1-\cos\phi_k^{(i)}\right) \biggr]
 \,. \nonumber
\end{align}

\subsection{Fluxons and quantum fluctuations in long Josephson junctions}

In an ideal undamped LJJ, the dynamics in the `long' $x$-direction
is governed by the Sine-Gordon equation (SGE),
\begin{equation}\label{eq:SGE}
 \partial_t^2\, \phi - c^2 \partial_x^2\, \phi  + \omega_J^2 \sin\phi = 0 \,,
\end{equation}
describing the LJJ phase field $\phi(x,t)$.
Here, $\omega_J = \sqrt{2\pi I_c/(\Phi_0 C_J)}$ is the LJJ plasma frequency,
$c=\lambda_J \omega_J$ is the Swihart velocity.
The SGE supports ballistic solitons (fluxons) as its fundamental solitonic
solution \cite{BaronePaterno, BaroneETAL1971},
which is characterized by the phase profile
\begin{eqnarray}\label{eq:soliton}
 \phi^{(\sigma, X)}(x,t)
 &=& 4 \arctan \exp\left(-\sigma \left(x - X(t)\right) \big/ W \right)
\,,
\end{eqnarray}
where $X$ is the center position of the fluxon,
the parameter $W = \lambda_J \sqrt{1 - v_0^2/c^2}$ is its characteristic width,
and $v_0 = \dot X$ its constant velocity ($|v_0| < c$).
The energy of the moving fluxon is $\Efl = 8 E_0/\sqrt{1 - v_0^2/c^2}$
where $E_0 = I_c \Phi_0 \lambda_J/(2\pi a)$ is the characteristic energy scale.
The fluxon polarity is set to $\sigma = 1$ in this work,
and we note that the opposite polarity ($\sigma = -1$) would yield identical readout results
if the external flux bias of the fluxonium were also inverted, $\phiext \to -\phiext$.

In the discrete array, the Josephson penetration depth
can be expressed as $\lambda_J = \sqrt{\Phi_0 a^2/(2\pi I_c L)} = a \sqrt{L_J/L}$,
where $a$ is the length of an LJJ unit cell.
As has recently been experimentally studied \cite{HanLJJ1},
despite finite discreteness $a/\lambda_J>0$ fluxons can move ballistically over large distances
$\gg \lambda_J$, i.e., with small energy loss \cite{WusOsb2020_RFL, BraKiv1998},
if $a/\lambda_J \ll 1$ and the initial velocity $v_0 < c$.

In an LJJ, the size of the quantum fluctuations is determined by the
parameter $\beta^2 = (4e^2/\hbar) \sqrt{L/C_J}$,
which is the ratio of the plasma wave impedance $\sqrt{L/C_J}$
and the resistance quantum $h/(2e)^2$.
Using $\beta^2$, the commutation relations of the phase fields in the LJJ
take the form $[\phi(x), \omega_J^{-1} \dot \phi(x')] = \beta^2 \lambda_J \delta(x-x')$.
These can be obtained in the continuum limit $\phi_k \to \phi(x_k)$
from the commutation relations $[\phi_k, n_l] = \ui \delta_{k,l}$
of the individual JJ phases $\phi_k$ in the LJJ
with the Cooper-pair number operator $n_k = \hbar/(2e)^2 C_J \dot{\phi}_{k}$.
Although $\beta^2$ is typically small,
tunneling of trapped fluxons over short distances has been observed \cite{Ustinov_Nature2003}.
Moreover, high kinetic-inductance films \cite{HKILJJ, TiN1, TiN2, TiN3}, including a foundry inductor \cite{SEEQCfab}, will allow the proposed high-impedance LJJs to be created.
We estimate that a discrete LJJ could be fabricated with $\beta^2 \approx 0.4$,
using a critical current density of $j_c = 0.1\,\mu\text{A}/\mu\text{m}^2$,
JJ areas of $0.23\,\mu \text{m}^2$, and high kinetic inductors,
and we use this value of $\beta^2$ in the present study (see App.~\ref{sec:specificfab}).
This LJJ can be made from a single mergemon-type JJ\cite{Mergemon} that uses a low critical current density from an annealed Nb JJ \cite{SchusterFab1, SchusterFab2}. Furthermore, for these specific fabrication parameters, the JJ frequency is $12.1 \uGHz$, such that it is easy for our fluxon in the LJJ to undergo the specified readout dynamics (near the qubit) in a time less than a $\uns$. This fabrication condition implies larger quantum fluctuations than previously considered for readout in Ref.~\onlinecite{AveRabSem2006}, but since $\beta^2 < 8\pi$
the solitons are still stable \cite{Coleman1975}. 

\subsection{Parameters for readout of qubit between two LJJs}
\label{sec:parameters}

In this study, we concentrate on a particular parameter case for the fluxonium and the LJJs to show that a single-fluxon readout of a qubit can be achieved in simulation.
The parameters are given in the caption of Fig.~\ref{fig1}.
Our simulations are performed in dimensionless units, scaled by the
characteristic LJJ parameters $L, I_c, C_J$,
which can individually depend on fabrication parameters.
In these units, the LJJ Lagrangian
only depends on the relative discreteness $a/\lambda_J = \sqrt{L/L_J}$,
and for negligible discreteness, we design at $L/L_J = 1/7$,
similar to our classical digital gate studies \cite{HanLJJ1}.
The interface and fluxonium parameters are also given in units of the LJJ characteristics.
Similar to digital gate interfaces \cite{WusOsb2020_RFL, WusOsb2023_BSR},
we set the LJJ-termination JJs and the rail-JJ of the interface to be slightly larger
and much larger than the JJs in the LJJ, respectively.
Additionally, these JJs have large shunt capacitances such that their plasma frequency is lowered relative to the JJs in the LJJ.
The chosen parameters enable scattering at the interface that results in conditional transmission for single-shot fluxon-based readout, as discussed in the following sections.
To find the scattering phenomena of interest,
we have simulated the fluxon dynamics over a range of qubit and interface-cell parameters.

In the fluxonium that we will present in the main text,
the energy scales satisfy the relations $E_J^q > 4 E_c^q \gg E_L^q$
(cf. Refs.~\onlinecite{Manucharyan2009},~\onlinecite{Manucharyan2019}).
Here we choose the fluxonium to approximately satisfy similar energy ratios
as in Ref.~\onlinecite{Manucharyan2009}:
$E_J^q/E_c^q \approx 5$ and $E_L^q/E_c^q \approx 0.2$.
With respect to the characteristic energy scale $E_0$ of the LJJ,
where $8 E_0$ is the rest energy of the fluxon,
the fluxonium energies are
$E_J^q/E_0 = \IJq a/(I_c \lambda_J)$,
$E_c^q/E_0 = \beta^4 C_J\lambda_J/(8 \CJq a)$, 
and $E_L^q/E_0 = L \lambda_J/(\Lq a)$. 
The transition energy of the first two fluxonium states of Fig.~\ref{fig1}
is $\hbar \omega_{01} \approx 2.4 E_c^q \approx 0.17 E_0$.
Assuming the specific LJJ characteristics of App.~\ref{sec:specificfab},
this value evaluates to $\omega_{01}/2\pi = 5.16 \uGHz$.
In comparison, the fluxon rest energy $8E_0$ is a factor $47$ larger,
and yet the qubit energy $\hbar\omega_{01}$ is sufficient to influence the system dynamics in the LJJs.

\begin{figure*}[t]
\includegraphics[width=\textwidth]{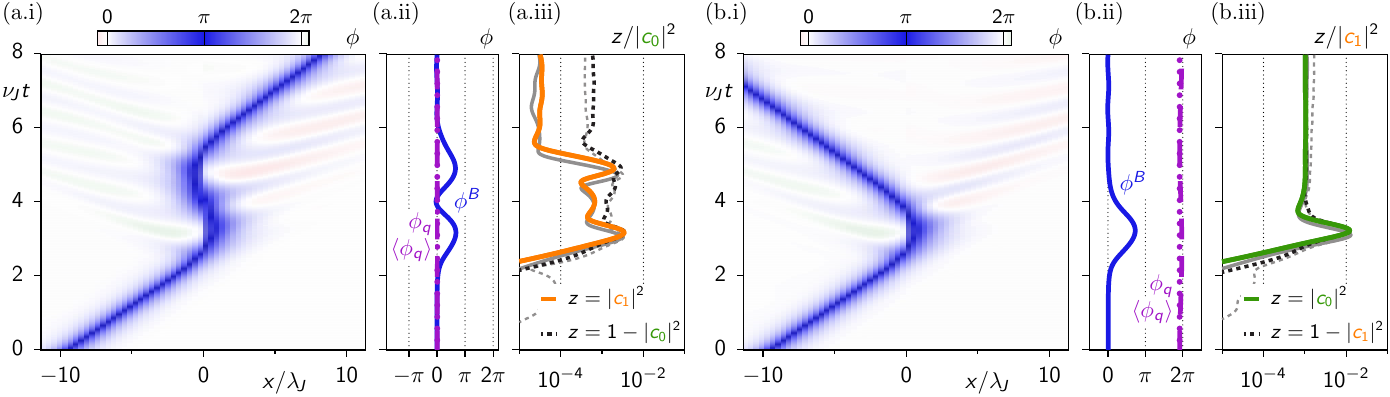}
 \caption{
 Simulated classical fluxon scattering at interface coupled to fluxonium,
 for fluxonium parameters of Fig.~\ref{fig1} and interface parameters
 $C_J^B/C_J = 11$, $I_c^B/I_c \approx 5.9$, $\hat{C}_J/C_J \approx 2.3$,
 and $\hat{I}_c/I_c \approx 1.3$.
 The fluxonium phase $\phiq$ is initialized as a classical variable,
 in the well of the fluxonium potential corresponding to
 the qubit state (a) $n=0$ and (b) $n=1$.
 Panels (i) show the dynamics of the JJ phases $\phi_k$ at positions
 in the left ($x_k < 0$) and right ($x_k > 0$) LJJ, respectively.
 Panels (ii) shows the classical evolution of the fluxonium phase $\phiq(t)$ (dashed)
 and rail-JJ phase $\phiBB(t)$ (solid).
 Using the data of $\phiBB(t)$ as an external drive to the Hamiltonian $H_q(t):=H_q(\phiq;\phiBB(t))$ (\Eq{eq:Hq1_phiq_phiBBbiased}),
 we calculate the time-evolution of the wave function $\psi(t)$ of the driven fluxonium and
 the overlap coefficients $c_m(t) = \langle \psi(t) | m(t) \rangle$
 with the instantaneous eigenstates $|m(t)\rangle$ of $H_q(t)$.
 Panels (iii) show the weights {\mycolor (infidelities)} of the fluxonium states that were not initially excited
 (orange or green solid and black dashed).
 Comparable values of the weights $|c_m(t)|^2$ (gray solid and gray dashed, respectively)
 were also obtained from the mixed quantum-classical dynamics of a collective-coordinate model, see Sec.~\ref{sec:mixedqucl}.
 From the final values of $\sum_{m\neq n}|c_{m}(t)|^2 /|c_{n}(t)|^2  = (1 - |c_{n}(t)|^2)/|c_{n}(t)|^2 \ll 1$
 we infer a total backaction of the readout on the qubit of $\approx 0.1\%$.
 In both initial-state cases (a,b), $\phiBB$ almost reaches the avoided crossing
 between the fluxonium states $n=0,1$ at $\phiBB = 0.8\pi$, cf. Fig.~\ref{fig1}(d),
 however, no significant inter-well tunneling occurs {\mycolor and the
  qubit state remained concentrated in the initial (excited) state}.
 For example, $\max(\phiBB(t)) = 0.73\pi$ in case (b) at $\nu_J t \approx 3.5$
 and at this time $\sum_{m \neq 1} |c_m(t)|^2 \approx |c_{0}(t)|^2 = 0.012 |c_{1}(t)|^2$.
 The expectation value
 $\langle \phiq \rangle(t) := \langle \psi(t) | \phiq | \psi(t) \rangle$
 (dotted) is also shown in panels (ii),
 confirming $|\langle \phiq \rangle(t)-\langle \phiq \rangle(0)|\ll 2\pi$
 due to neglible tunneling.
 The good correspondence between $\langle \phiq \rangle(t)$
 and $\phiq(t)$ (dashed) suggests that the classical approximation is reliable. Case (a) (n=0) exhibits two bounces in the time duration of 3 Josephson periods, and for the achievable period of $2\pi/\omega_J$=1/12.1 $\uGHz$, the duration is $\ll$ 1 $\uns$.
 }
 \label{fig2}
\end{figure*}

From \Eq{eq:Lagr_intf} we see that the interaction energy
between the fluxonium and the interface is determined by $E_L^q$, cf.~\Eq{eq:Vint}.
Since the dynamics of the fluxon, on the other hand, occurs with the energy scale of $E_0$,
an impact of the qubit on the moving fluxon requires that
$\pi^2 E_L^q/E_0 = \pi^2 (E_L^q/E_c^q)\cdot(E_c^q/E_0)$ is small but not too small compared to 1.
It then follows, since $E_L^q/E_c^q < 1$ in the fluxonium,
that $E_c^q/E_0 = \beta^4 C_J \lambda_J/(8\CJq a)$ needs to be sufficiently large
and therefore that $\beta^2$, the scale of quantum fluctuations in the LJJ,
is not too small.
Similar to the estimate above, here we use $\beta^2 = 0.4$.

Lastly, the strongly coupled qubit-interface system
needs to avoid a regime where quantum fluctuations in the direction of $\phiBB=\phi_{L}-\phi_{R}$ are larger than those in $\phiq$-direction.
In that case, the ground and first excited states of the biased hybridized system would lie in the same well such that the flux states are indistinguishable.
In leading approximation, the harmonic confinement in these directions is
given by the corresponding plasma frequencies
$\omega_J^x = (2\pi I_c^x/(\Phi_0 C_J^x))^{1/2}$ (see App.~\ref{app:system}),
such that the incompatible regime corresponds to
$\omega_J^q \gg \omega_J^B$.
In our mainly studied case, the JJ frequency ratios are
$\omega_J^B/\omega_J = \sqrt{\IJbb C_J/(I_c \CJbb)} = 0.73$
and $\omega_J^q/\omega_J = \sqrt{\IJq C_J/(I_c \CJq)} = 1.2$, and thus they lie outside of the incompatible regime. These values imply that the interface JJs need to be capacitively shunted relative to the LJJ. Additionally, the JJs in the LJJ need to be shunted or the qubit JJ needs to be fabricated separately.

Based on our observations in digital gate interfaces \cite{WusOsb2020_RFL, WusOsb2023_BSR},
we expect the interface JJ phase $\phiBB$ to undergo significant changes
when the fluxon arrives at the interface.
As seen in the last line of \Eq{eq:Lagr_intf},
$\phiBB$ biases the fluxonium in the same way as the external flux $\phiext$.
Since the fluxonium after the readout should be in the same configuration as before,
we only consider computational cases of the scattering outcomes
where $\phiBB$ returns to its initial value after the scattering.
In order to minimize state mixing and excitation
of higher-lying fluxonium states during the readout,
the $\phiBB$-amplitude must remain limited during the fluxon scattering such
that the fluxonium is not driven through avoided crossings.
For the same reason, we configure the fluxonium away from the half-flux quantum sweet spot,
choosing $\phiext = 0.2\pi$,
see Fig.~\ref{fig1}(d).
This choice allows both for a large maximum amplitude of $\phiBB \lesssim 0.8\pi$
to the avoided crossing at the half-flux sweet spot $\phiext = \pi$,
as well as for some degree of residual fluctuations $|\phiBB| < 0.2\pi$ to the avoided crossing at $\phiext = 0$.
The intended readout therefore falls into the regime (a) studied in Ref.~\onlinecite{FedorovETAL2007},
however that work studied the time-delay readout mode, and we are reporting on the transmission readout mode.

Although quantum fluctuations in the LJJs themselves are small, in principle
the quantum behavior of a fluxon may be characterized by a wave function~\cite{AveRabSem2006}
$\Psi(X)$ as function of the fluxon position $X$.
For low discreteness effects,
one may impose the condition $\lambda_{\text{dB}} > a$ on its associated de-Broglie wavelength,
similar to the classical condition $\lambda_J > a$.
From this criterion, we obtain an upper bound on the fluxon velocity, cf.~Sec.~\ref{sec:discussion},
which amounts to $v/c < 0.8$ for our chosen parameters $\beta^2=0.4$ and $a/\lambda_J = 1/\sqrt{7}$.
Similar to our classical gate simulations, here we use the initial fluxon velocity of $v/c=0.6$,
to satisfy the condition.

While the quantum dynamics of a single LJJ may be characterized by the single wave function $\Psi(X)$,
we are studying a qubit strongly coupled to the interface between two LJJs, and we are interested in how the qubit state affects the fluxon scattering at this interface.
Such a process is better described in terms of
(up to normalization) independent wave functions for each of the LJJs, $\Psi_l(X<0)$ and $\Psi_r(X>0)$.
This is related to a model which uses two
independent coordinates $X_{L,R}$ in the left and right LJJ.
The classical and quantum dynamics of this model are discussed below
in Secs.~\ref{sec:CCM} and \ref{sec:mixedqucl}, respectively.

\subsection{System state prior to fluxon scattering}
\label{sec:initialstate}

Let us first consider the system in the absence of a fluxon
in the LJJs.
Due to strong coupling with the interface JJs,
one can expect that a finite excitation of the qubit phase $\phiq$
gives rise to finite phases $\phiBB$ and $\phiL,\phiR$.
The latter two, in turn, will generate evanescent fields in the LJJs,
localized at the interface.
In general, these LJJ excitations have to be taken into account
to describe the steady state of the circuit,
as outlined in App.~\ref{app:system}.
For our particular parameter case, however,
since $E_L^q, E_J^q \ll E_J^B$, the amplitudes satisfy $|\phiBB| \ll |\phiq|$,
and $\phi_{L,R} \ll |\phiq|$.
Figure \ref{fig1}(c) illustrates the wave functions
$\langle \phiq, \phiBB | \psi \rangle_n$
of the three lowest eigenstates
of the quantum Hamiltonian $H^{(2)}(\phiq, \phiBB)$, \Eq{eq:H2_boundstate},
for the coupled fluxonium-interface system,
which accounts for the evanescent LJJ excitations.
From the eigenstates of the coupled system, we observe that the quantum fluctuation on the interface JJ is small compared with that
of the fluxonium,
$\langle (\Delta \phiBB)^2 \rangle \ll \langle (\Delta \phiq)^2 \rangle$.
One may therefore approximate the fluxonium as initially decoupled from
the interface.
Accordingly, we set $\phiBB = \phi_{L,R} = 0$ in the initial state of the simulations.

\section{Classical scattering dynamics}\label{sec:classical}

Considering the classical nature
of the fluxon in the LJJ (small quantum fluctuations $\beta^2\ll 1$)
and the macroscopically distinquishable qubit states,
the readout process may be described in a semi-classical approximation.
For the chosen parameters, this condition is fulfilled
with the two qubit states $n=0,1$ localized in separate wells of the fluxonium potential,
at $\phiq \approx 0$ and $\phiq \approx 2\pi$, respectively, see Fig.~\ref{fig1}(b,c).

\subsection{Full circuit simulation}\label{sec:circuitsimulation}

To obtain the semiclassical scattering dynamics we simulate the circuit
equations of motion governed by the Langrangian \Eq{eq:Lagrangian}.
The initial state of the simulation consists of
a ballistic (right-moving) fluxon in the left LJJ, modeled with \Eq{eq:soliton},
while the fluxonium phase is set according to
the expectation values
$\phiq(t=0) = \langle \phiq \rangle_{n} = \langle n | \phiq | n \rangle$
in one of the quantum states $n=0,1$ of $H_q$, \Eq{eq:Hq_phiq}.
More generally, the
quantum states $n$ of the coupled fluxonium-LJJ Hamiltonian, \Eq{eq:H2_boundstate},
could be used such that in addition to $\phiq(t=0)$,
$\phiBB(t=0) = \langle \phiBB \rangle_{n} = \langle n | \phiBB | n \rangle$
is also determined (with corresponding evanescent phases in the LJJs).
However, this refinement yields similar results for the parameters used here.

Figure ~\ref{fig2} shows the simulation results, for the appropriate fluxonium initial-state values: (a) $\phiq = 0.0095\pi$ for $n=0$ and (b) $\phiq =1.917 \pi$ for $n=1$.
Both cases are illustrated in panels (i) by the LJJ-junction phases $\phi_k(t)$ shown
as functions of the position $x_k = \mp a (k + 1/2) \lessgtr 0$ ($k=0,1,2,\ldots$)
in the left and right LJJ, respectively.
Additionally, in panels (ii) the evolution is shown
for the phases $\phiBB(t)$ (blue solid)
and $\phiq(t)$ (purple dashed).

In Fig.~\ref{fig2}(a) the fluxon is seen to scatter forward from the input to the output LJJ, with a couple of bounces
-- with a bounce to backward motion and then to forward motion
-- as opposed to a direct transmission process \cite{SpectralWalls}.
Panel (ii) shows that the interface rail-JJ assumes a relatively large amplitude
in $\phiBB \lesssim \pi$ during the bounces,
again in contrast to direct fluxon transmission where the amplitude is generally much smaller, cf.~Fig.~3 in Ref.~\onlinecite{WusOsb2023_BSR}.
In Fig.~\ref{fig2}(b) the initial dynamics of the fluxon at the interface is simpler, where the fluxon bounces (scatters) backward into the input LJJ.
In both cases (a,b), the classically evolving fluxonium phase $\phiq(t)$ remains close to its initial value.
This is a necessary (but not sufficient) condition for the validity
of the semiclassical approximation.
In contrast, if $\phiq(t)$ would evolve away from $\phiq(0)$, the initial semiclassical approximation, $\phiq \approx \langle \phiq \rangle$, would no longer be fulfilled.

The fluxon shape changes significantly during scattering,
and, in particular, temporarily breaks open at the interface.
The break can be seen as a relatively large amplitude of the phase difference across the interface cell, $\phiL - \phiR = \phiBB \gtrsim 0.7\pi$
observed during the scattering.
This value is much larger than the maximum phase difference across a cell
in our LJJ bulk cells
(estimated as $\text{max}(\phi_k-\phi_{k+1}) = 2 a/(\sqrt{1-(v/c)^2} \lambda_J)$,
and amounts to $0.3\pi$ for our $\lambda_J/a=\sqrt{7}$ and initial velocity $v/c=0.6$).
The breaking of the fluxon into two parts at the interface is enabled
by the rail JJ of the interface, as it compensates for the resulting phase difference
between the left and right LJJ.
This process is comparable to the formation of `semifluxons'
in LJJs with embedded $0-\pi$ JJs \cite{Goldobin2002}.

The distinct dynamics for the different fluxonium states
shows a transmission-mode readout of the qubit state using a single fluxon --- the fluxon is scattered forward in one state, while it is reflected in the other.

{\mycolor
While Fig.~\ref{fig2} illustrates the simulation results when the initial fluxonium phase is set to the expectation values $\phiq(t=0) = \langle \phiq \rangle_n$, we have checked that the same essential dynamics occurs under small variations $|\Delta \phiq(t=0)| \leq 0.4 \pi$ of this initial phase. 
As this exceeds the quantum uncertainties $\sqrt{\langle (\Delta \phiq)^2 \rangle} \approx 0.28 \pi$ of the first two fluxonium states, it demonstrates that our classical simulation is consistent with expected qubit fluctuations. 
Moreover, it also indicates robustness of the readout against these fluctuations.
}

\subsection{Collective coordinate model and simulations}
\label{sec:CCM}

\begin{figure*}[t]
\includegraphics[width=\textwidth]{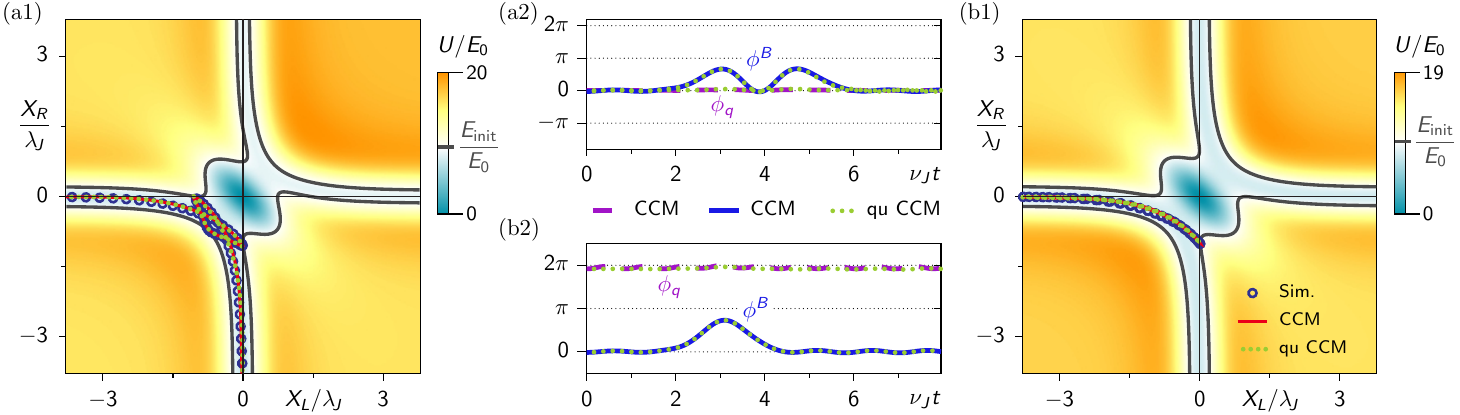}
 \caption{
The CC potentials $U(X_L,X_R,\phiq)$, \Eq{eq:U_CC_cl},
 with $\phiq$ fixed at (a) $\langle \phiq \rangle_{n=0} = 0.0095\pi$
 and (b) $\langle \phiq \rangle_{n=1} = 1.917 \pi$, respectively.
 Equipotential lines (gray) are shown at the initial energy
 $E_{\text{init}} = \Efl + U_q(\langle \phiq \rangle_{n})$.
 Panels (a1,b1) show the components $(X_L,X_R)(t)$ and panels (a2,b2) show
 the component $\phiq(t)$ of the trajectories $(X_L,X_R,\phiq,\phiBB)(t)$,
 which are calculated with three different methods:
 (i) the classical CC equations of motion,
 \Eqs{eq:EOM0_CC} and \eqref{eq:EOM1_CC} with the potential \eqref{eq:U_CC_cl},
 (red solid line in (a1,b1) and purple dashed in (a2,b2)),
 (ii) the mixed quantum-classical equation of motion,
 \Eqs{eq:EOM0_CC} and \eqref{eq:Schroedinger_mixedqucl}
 with the potential \eqref{eq:U_CC_mixedqucl} (green dotted lines),
 and (iii) the results $\phi_n^{(l,r)}(t)$ of the full circuit simulation,
 fitted to the form of \Eq{eq:fluxoncombination} (blue markers in panels (a1,b1)).
 Panels (a2,b2) also show $\phiBB=\phiL-\phiR$ for the first two methods ($\phiBB$
 for the third method is shown in Fig.~\ref{fig2}).
 }
 \label{fig3}
\end{figure*}

{\mycolor
A collective coordinate model is a standard approximation method for highly nonlinear systems with soliton-like solutions, such as for a soliton in a sine-Gordon system \cite{McLaughlinScott1978, Rajaraman}.
}
Fluxon scattering at circuit interfaces,
similar to that of Fig.~\ref{fig1}(a),
is well described by a collective-coordinate (CC) model,
where the fields in the left and right LJJ are each modeled as a superposition of fluxon and mirror antifluxon,
\begin{align}
\label{eq:fluxoncombination}
\phi(x<0) &=
   \phi^{(\sigma, X_L)}(x) + \phi^{(-\sigma, -X_L)}(x) - 2\pi (1-\sigma)
   \\
\phi(x>0) &=
   \phi^{(-\sigma, X_R)}(x) + \phi^{(\sigma, -X_R)}(x) - 2\pi
 \,. \nonumber
\end{align}
This builds upon previous reversible-digital studies \cite{WusOsb2020_RFL, WusOsb2023_BSR}, 
{\mycolor 
where CC models were found to accurately describe e.g.~the dynamics of reversible-digital gates with high energy conservation. 
}
In \Eq{eq:fluxoncombination}, $\phi^{(\sigma, X)}$ is defined by the soliton solution of the continuous LJJ
field, \Eq{eq:soliton}.
The time-dependent fluxon positions $X_{L,R}(t)$ serve as the dynamical
coordinates of the model, while the fluxon width $W$ is
taken to be constant in a so-called adiabatic approximation \cite{DauxoisPeyrard}.
The ansatz \eqref{eq:fluxoncombination}
conveniently parametrizes all possible asymptotic states
to which a single fluxon of polarity $\sigma$
can scatter elastically after moving towards the interface from the left LJJ.
Asymptotically, if the coordinate $X_i$ ($i=L,R$) is far away from the interface,
the field in the LJJ $i$, defined in \Eq{eq:fluxoncombination},
can be either fluxon- or antifluxon-like.
Furthermore, \Eq{eq:fluxoncombination} approximates localized excitations at the interface as observed in classical resonant fluxon scattering ~\cite{WusOsb2020_RFL, WusOsb2023_BSR}.

Equation \eqref{eq:fluxoncombination} does not take into account
possible evanescent excitations of the LJJ fields near the interface
(see \Eq{eq:ansatz_boundstate}).
However, due to the small amplitudes of these localized fields
in our circuit, this approximation seems justified.
The linear superposition of solitons in \Eq{eq:fluxoncombination}
also does not account for the full nonlinear nature of the LJJ-dynamics.
{\mycolor
We have found that the CC trajectories agree with the full circuit simulations for all of our classical (digital) ballistic gates to good accuracy. 
They have the property of being nearly energy conserving (by comparing the output to input fluxon energy ratio). 
For example, the same ansatz worked in a comparable shift register cell \cite{WusOsb2023_BSR}, which contains a storage inductor at the interface in place of the qubit present in this work. 
Below, we will use the ansatz for a tractable quantum treatment of the scattering dynamics, cf.~Sec.~\ref{sec:mixedqucl}.
}

Using \Eq{eq:fluxoncombination}, the many
JJ degrees of freedom in the two LJJs reduce to only two coupled degrees of freedom $(X_L, X_R)$.
As described in App.~\ref{app:CCM},
the remaining three coordinates $(X_L, X_R, \phiq)$
then obey the classical equations of motion
\begin{eqnarray}
\label{eq:EOM0_CC}
\!\!\!\!\! \left(\!\!\begin{array}{c} \ddot X_L \\[1ex] \ddot X_R \end{array} \!\!\right)
\!\!&=&\!\! - \mathbf{M}^{-1} \left(\!\!\begin{array}{l}
 c^2 \frac{\partial U}{\partial X_L}
 + \frac{1}{2}\frac{\partial m_L}{\partial X_L}  \dot X_L^2
 + \frac{\partial m_{LR}}{\partial X_R} \dot X_R^2 \\[1ex]
 c^2 \frac{\partial U}{\partial X_R}
 + \frac{1}{2}\frac{\partial m_R}{\partial X_R} \dot X_R^2
 + \frac{\partial m_{LR}}{\partial X_L} \dot X_L^2
 \end{array} \!\!\right) \\
 \label{eq:EOM1_CC}
 \ddot{\phi}_{q} &=& - \frac{1}{M_q} \frac{\partial U}{\partial \phiq}
 \,,
\end{eqnarray}
with the dimensionless potential
\begin{equation}\label{eq:U_CC_cl}
 U(X_L,X_R,\phiq) = \Ufl(X_L,X_R) + \frac{\Uq(\phiq)}{E_0} + \frac{\Vint(X_L,X_R,\phiq)}{E_0}
 \,,
\end{equation}
whose components are given in \Eq{eqA:Ufl_CCM}-\eqref{eqA:Vint_CCM}.
The mass matrix $\mathbf{M}$ in \Eq{eq:EOM0_CC} is composed
of the coordinate-dependent dimensionless elements $M_{ii} = m_i (X_i)$
and $M_{i,j\neq i} = m_{LR}(X_L, X_R)$ ($i, j = L, R$).
The position dependence of these elements gives rise to some
unusual terms in \Eq{eq:EOM0_CC}.
The fluxonium mass is here also expressed in LJJ units,
$M_q = C_J^q a/(C_J \lambda_J)$.

Panels (a1,b1) in Fig.~\ref{fig3} illustrates the potential
$U(X_L,X_R,\phiq)$, with $\phiq$
fixed at values $\phiq(t=0) = \langle \phiq \rangle_{n}$
for the two initial fluxonium states $n=0,1$.
The difference of $U(X_L,X_R,\phiq)$ between the two cases approximately scales with the energy $E_L^q$ of the fluxonium shunt inductance, while $U$ dominantly scales with $E_0$.
Since $E_L^q/E_0 \ll 1$ with the chosen parameters,
it is not surprising that the potential for the two initial states of the qubit look very similar.
In (a1) the mirror symmetry about the line $X_R = -X_L$ is more strongly broken
than in (b1),
as can be seen by looking at the intersection of the
lines $(X_L,X_R=0)$ and $(X_L=0,X_R)$ with the equipotential line $U = E_{\text{init}}$.
We note that other parameter regimes may cause more significant qualitative differences of $U$
between different interface states, cf.~Fig.~9 in Ref.~\onlinecite{WusOsb2023_BSR}.

The potential of Fig.~\ref{fig3} has a central well at $(X_L,X_R) = (0,0)$,
and four asymptotic scattering channels.
These correspond to a fluxon in the left LJJ ($X_L/\lambda_J < -1$, $X_R=0$) or in the right LJJ ($X_L=0$, $X_R/\lambda_J>1$),
and an antifluxon in the left LJJ ($X_L/\lambda_J > 1$, $X_R=0$)
or in the right LJJ ($X_L=0$, $X_R/\lambda_J > 1$), respectively.
Gray equipotential lines are shown
at the initial energy, $E_{\text{init}} = \Efl + U_q(\langle \phiq \rangle_n)$,
and indicate that the scattering channels are connected
and energetically allow the incoming fluxon to evolve towards any of these
scattering outcomes.
The potential thus allows in principle even polarity-inverting scattering,
as it is used for digital-reversible logic \cite{WusOsb2023_BSR, WusOsb2020_RFL},
However, this type of dynamics is not desirable for the qubit readout,
as will be discussed at the end of Sec.~\ref{sec:mixedqucl}.

From \Eq{eq:EOM0_CC} and initial state $X_L/\lambda_J \ll -1$, $X_R=0$,
and $\phiq = \langle \phiq \rangle_{n}$
we obtain CC trajectories $(X_L,X_R,\phiq)(t)$, shown as blue lines
in panels (a1,b1) of Fig.~\ref{fig3}, together with $\phiq$ in panels (a2,b2).
We also compare the CC trajectories with ``simulated'' trajectories, which are
obtained by fitting the phases $\phi_{k}^{(l,r)}(t)$ of the circuit simulations
to \Eq{eq:fluxoncombination} (blue markers).
As Fig.~\ref{fig3} shows,
there is good quantitative agreement between simulation and CC-model trajectories.

Despite the similarity of the CC potentials
$U(X_L,X_R,\phiq = \langle \phiq \rangle_{n})$,
the trajectories for the two fluxonium states shown in (a) and (b) differ qualitatively from each other.
For case (a),
with the fluxonium initialized in its ground state, the trajectory undergoes a sequence of three bounces off the CC potential --- an initial and final bounce that change the direction
and a middle bounce that scatters as a grazing incidence redirection.
From the bounces, the trajectory moves into the scattering channel at $X_R\ll0$,
corresponding to a fluxon in the right LJJ.
For case (b), however, the slight quantitative change to the CC potential
causes a single bounce off the potential
such that the trajectory gets back-reflected to $X_L\ll0$, corresponding to the left LJJ.

As mentioned before, during the scattering the fluxon partially breaks into two `semifluxons',
as a result of the strong scattering potential formed by
the interface (which is modified by the qubit).
This behavior cannot be modeled by fluxon perturbation theory
\cite{McLaughlinScott1978}, which assumes an unchanging fluxon shape
and parametrizes the fluxon with a single coordinate.
Fluxon perturbation theory has been used in previous fluxon-readout studies \cite{FedorovETAL2007, Ustinov2013,Kuzmin2015},
which is appropriate when weak perturbations are induced by the qubit on the fluxon's potential within a single LJJ.
In contrast, our CC model with two independent fluxon coordinates
approximates the interface scattering very well.

{\mycolor
Similar to the quantum consistency check of the classical circuit simulation (cf.~last paragraph of Sec.~\ref{sec:circuitsimulation}), we have checked that small deviations $|\Delta \phiq(t=0)| \leq 0.3\pi$ of the initial phase $\phiq(t=0)$ from  $\langle \phiq \rangle_n$ give rise to the same essential CC dynamics as seen in Fig.~\ref{fig3}.
}

\section{Quantum dynamics}\label{sec:quantum}

\subsection{Estimating backaction on qubit}
\label{sec:backaction}

Here we estimate the effect of the fluxon back on the qubit.
To this end, we consider the subsystem consisting of the fluxonium and the rail JJ of the interface,
with the Hamiltonian
\begin{align}
\label{eq:Hq1_phiq_phiBBbiased}
 H_q(\phiq;\phiBB) &= 4 E_c^q \nq^2 + E_J^q ( 1 - \cos \phiq ) \\
&\quad + \frac{E_L^q}{2} (\phiq - \phiBB - \phiext)^2 \,. \nonumber
\end{align}
In this analysis, we treat $\phiBB$ as an time-dependent external parameter that drives the fluxonium.
By taking $\phiBB(t)$ from the classical circuit
simulation as the drive (cf.~panels (ii) in Fig.~\ref{fig2}(b)),
we study the quantum evolution of a wave packet $\langle \phiq | \psi \rangle$,
subject to the time-dependent Hamiltonian $H_q(t) = H_q(\phiq;\phiBB(t))$.

We determine the eigenstates $| n \rangle$ and eigenenergies $\epsilon_n$
of \Eq{eq:Hq1_phiq_phiBBbiased} at $t=0$, where $\phiBB(t=0)=0$,
and initialize the wave packet as one of the two lowest states,
$|\psi(t=0)\rangle = | n \rangle$.
The wave packet $| \psi(t) \rangle$ is then evolved under the time-dependent
Hamiltonian \eqref{eq:Hq1_phiq_phiBBbiased}.
In this time-dependent quantum evolution, the wave packet remains localized
in its original well,
even when $\phiBB(t)$ approaches the avoided level crossing of fluxonium states $n=0,1$
at $\phiBB + \phiext = \pi$, cf. Fig.~\ref{fig1}(d).
The absence of quantum tunneling is also seen in the expectation value
\begin{equation}\label{eq:ephiq_phiBBdrivenfluxonium}
 \langle \phiq(t) \rangle = \langle \psi(t)| \phiq |\psi(t) \rangle \,,
\end{equation}
shown in panels (ii) of Fig.~\ref{fig2} (dotted line),
as it remains close to its initial value throughout,
in good agreement with $\phiq(t)$ obtained from the classical circuit
simulation (dashed).

For each moment in time $t$ we also determine the instantaneous eigenstates $| m(t) \rangle$
and eigenenergies $\epsilon_m(t)$ of \Eq{eq:Hq1_phiq_phiBBbiased},
at bias value $\phiBB(t)$.
Expanding the evolved wave function $\psi(t)$ in this basis,
\begin{equation}\label{eq:cn_phiBBdrivenfluxonium}
 c_m(t) := \langle \psi(t) | m(t) \rangle
 \,,
\end{equation}
we quantify the perturbation of the qubit by the drive.
Panels (iii) in Fig.~\ref{fig2} show the occupation
$\sum_{m \neq n} |c_{m}(t)|^2 = 1 - |c_{n}(t)|^2$
in all states other than the initial state $n$ (black dashed lines).
With $\phiBB(t)$ approaching the avoided level 
at $\phiBB \approx 0.8$,
the partner state of the avoided crossing (solid color line)
naturally starts to be occupied.
However, this occupation remains small, reaching a maximum of $\approx 1 \%$
e.g. for (b) $n=1$, where $\max(\phiBB) = 0.73\pi$ at $\nu_J t \approx 3.5$
(at this time $\sum_{m \neq 1} |c_m(t)|^2 \approx |c_{0}(t)|^2 = 0.012 |c_{1}(t)|^2$).
The weight remains concentrated in the initially excited state,
and no significant interwell-tunneling takes place.
This shows that the qubit state remains minimally perturbed by the drive.

Similarly, in a (static) two-level approximation of the avoided crossing,
the upper eigenvector $\mathbf{v}_{\text{up}} = (v_{m=0}, v_{m=1})$
(expressed in the diabatic basis) has only
$|v_{m=0}|^2 = |\langle m=0 | v_{\text{up}} \rangle|^2 = 0.016$
at this (static) $\phiBB$-bias.

After fluxon scattering, the occupations of all states except the initially excited state $n$
sums up to $\sum_{m \neq n} |c_{m(t)}(t)|^2 \approx 10^{-3}$
in both cases (a,b), i.e. the backaction of the readout on the qubit is $\approx 0.1\%$.

\subsection{Mixed quantum-classical dynamics}
\label{sec:mixedqucl}

The SGE, \Eq{eq:SGE}, in an ideal, infinite LJJ is known to be integrable
and thus can be quantized in a non-perturbative way using methods of
quantum field theory \cite{Fadeev1978}.
However, the integrability is broken in our LJJ-system,
comprised of two LJJs and a qubit, all connected by an interface cell.
A general quantum dynamical treatment of the full system is difficult
due to the many JJs (degrees of freedom) in the underlying circuit model,
\Eq{eq:Lagrangian}.
However, the CC model described in Sec.~\ref{sec:CCM}
has only three degrees of freedom.
At this reduced level of complexity, the numerical time evolution of the
quantized system is tractable.

From the Lagrangian in the CC approximation, \Eq{eqA:Lagr_CCM},
we obtain the corresponding Hamiltonian,
\begin{align}\label{eq:Ham_CC}
H &= H_{\text{fl}}(X_L,X_R) + H_q(\phiq) + \Vint(X_L,X_R,\phiq) \\
H_{\text{fl}} &=
\frac{c^2}{2 E_0} \begin{pmatrix} P_L \\ P_R \end{pmatrix}
\mathbf{M}^{-1} \begin{pmatrix} P_L \\ P_R \end{pmatrix}
+ \Ufl(X_L,X_R)
\end{align}
with $H_q$, $\Vint$, and $\Ufl$ from \Eqs{eq:Hq_phiq}, \eqref{eq:Vint},
and \eqref{eqA:Ufl_CCM},
and with the CC momenta 
\begin{equation}\label{eq:CC_momenta}
 \begin{pmatrix} P_L \\ P_R \end{pmatrix}
 = \frac{E_0}{c^2}\; \mathbf{M}
 \begin{pmatrix} \dot{X}_L \\ \dot{X}_R \end{pmatrix}
 \,.
\end{equation}

The masses for the qubit and fluxon degrees of freedom in \Eq{eq:Ham_CC}
fulfill $M_q \ll m_{L,R}$,
where $M_q \approx 0.3$  and $m_{L,R} \geq 8$
(e.g. $m_L$ ranging between $m_0 = 8\lambda_J/W$  at $|X_L|\gg \lambda_J$
and $16\lambda_J/W + 12.4 a/\lambda_J$ at $X_L=0$)
and we will use this large mass imbalance to treat the light and heavy
degrees of freedom separately.
This situation is comparable to molecular systems,
where the vastly larger mass of the nuclei relative to the electron mass
allows one to treat the motion of the former classically
while computing the quantum dynamics of the electronic degrees of freedom.
In the so-called adiabatic approximation,
the electronic degrees of freedom are assumed for all times to be
in the ground state of the potential, which is determined by the slowly-evolving
nuclei positions \cite{commentBornOppenheimer}.

Here we adopt a similar approach,
assuming classical dynamics for the heavy coordinates in the LJJs, $X_{L}$ and $X_{R}$,
and quantum dynamics for the light qubit degree of freedom.
However, we do not assume the latter to evolve adiabatically.
The classical equations of motion for the coordinates $X_{L,R}$ are
of the same form as \Eq{eq:EOM0_CC}, but now with the dimensionless potential
\begin{equation}\label{eq:U_CC_mixedqucl}
U(X_L,X_R) = \Ufl(X_L,X_R) + E_0^{-1} \langle \psi(t) | \Vint | \psi(t) \rangle
\end{equation}
replacing \Eq{eq:U_CC_cl}.
Here, $\langle \phiq |\psi(t)\rangle$ is the qubit wave function
following the Schr\"odinger time evolution
\begin{equation}\label{eq:Schroedinger_mixedqucl}
 -\ui \hbar \frac{d}{d t} |\psi(t)\rangle
   = \left( H_q(\phiq) + \Vint \right) |\psi(t)\rangle
 \;,
\end{equation}
where $\Vint$ parametrically depends on the current values of
the CC coordinates $X_{L,R}(t)$.
In contrast to the fully classical CC model, \Eqs{eq:EOM0_CC}--\eqref{eq:U_CC_cl},
where all three coordinates $(X_{L},X_{R},\phiq)$
are classical, $\phiq$ is here treated as a quantum variable.
From the solution of \Eq{eq:Schroedinger_mixedqucl} at each time $t$ 
we obtain the expectation values
$\langle \phiq \rangle = \langle \psi(t) | \phiq | \psi(t) \rangle$
and with these we can evaluate the forces 
exerted by the qubit on the fluxon coordinates $X_{L,R}$ in \Eq{eq:EOM0_CC},
\begin{equation}
 \frac{\partial}{\partial X_i} \langle \psi(t) | \Vint | \psi(t) \rangle
 = \mp E_L^q \frac{\partial \phi_i}{\partial X_i}  \left(\langle \phiq \rangle - \phiext \right)
 \,,
\end{equation}
where $i=L,R$ and the upper (lower) sign are for $i=L$ ($i=R$), respectively.

Initially, when the fluxon is far away from the interface, $\Vint=0$,
and we initialize $|\psi(t=0)\rangle$ as the $n$-th eigenstate of the isolated fluxonium qubit, $H_q$.
As discussed in Sec.~\ref{sec:initialstate},
this serves as a reasonable approximation of the quantum bound state $n$.

In Fig.~\ref{fig3},
the evolution of the mixed quantum-classical approximation (dotted)
are compared with the purely classical evolution (solid and dashed) of the CC model.
For both initial qubit states $n=0,1$,
there is excellent agreement between these two approaches.
Throughout the dynamics of the CC coordinate $X_{L,R}$, the
effective qubit potential
phase difference $\phiBB=\phiL-\phiR$ remains constrained, such that
the effective qubit potential $U_q + \Vint$ is never tilted far enough
to induce tunneling of the qubit wave function from one well to the next.
As a result, $\langle \phiq \rangle$ remains close to its initial value.

Similar to Sec.~\ref{sec:backaction},
we can also determine the instantaneous eigenstates $|m(t)\rangle$
and eigenenergies $\varepsilon_m(t)$ of the Hamiltonian
in \Eq{eq:Schroedinger_mixedqucl}, and expand the evolving wave function
$|\psi(t)\rangle$ in this basis, cf.~\Eq{eq:cn_phiBBdrivenfluxonium}.
In Panels (iii) of Fig.~\ref{fig2},
the resulting occupations $\sum_{m\neq n} |c_{m}(t)|^2$ in all states not equal to the initial state $n$ are shown as gray dashed lines.
They agree well with the values obtained from the analysis of
Sec.~\ref{sec:backaction}.

For the presented parameter regime,
we also find good agreement with the evolution
according to the above-mentioned adiabatic approximation,
where the qubit is assumed to always remain
in the quantum state $n$ in which it had been initialized.
Note that in the adiabatic approximation, \Eq{eq:Schroedinger_mixedqucl} is replaced
by the corresponding time-independent Schr\"odinger equation,
giving the eigenstate $|\psi_n(t)\rangle$ and the resulting expectation value $\langle \psi_n(t) | \Vint | \psi_n(t) \rangle$, which is inserted into \Eq{eq:U_CC_mixedqucl} at each time step.
In fact, the adiabatic assumption
is well justified in the situation of Fig.~\ref{fig2},
as seen in the analysis of the previous section.

As an unsuitable readout example,
we briefly mention a parameter case near the half-flux point, $\phiext = 0.8\pi$,
which we have also studied.
In this case, $\phiBB$ undergoes large-amplitude changes during the dynamics,
for either of the initial qubit states $n$.
For $n=1$, the classical scattering from input to output LJJ induces inversion of the fluxon polarity
which involves a permanent $4\pi$-change of $\phiBB$,
similar to a ballistic NOT gate \cite{WusOsb2020_RFL}.
The rate $\text{d} \phiBB/\text{d} t$ is sufficiently large that the
fluxonium undergoes near-perfect diabatic transitions through various avoided crossings
while remaining confined to one well,
and ends up in state $n=4$.
Since no interwell-tunneling is involved, we find good agreement between
the fully classical and the mixed quantum-classical dynamics.
In contrast, for $n=0$ the classical dynamics predicts backscattering of the
fluxon after reaching the output LJJ
(a NOT followed by a time-reversed NOT, comprising two scattering events). Within the classical CC model, this is explained
by the asymmetry in the CC potential at $\langle \phiq \rangle \approx 0$.
It leads to a closure of the scattering channel which corresponds to
forward scattering of a fluxon with polarity inversion (NOT channel).
However, the rate $\text{d} \phiBB/\text{d} t$ during the first interface scattering is
sufficiently large such that all Landau-Zener transitions
are diabatic until the fluxonium is excited to high-lying states $n\approx 15$,
and then distributed among several states.
As a result, $\langle \phi_q \rangle$ deviates from $0$,
the NOT channel opens in the potential, and the CC trajectory $X_{L,R}(t)$
evolves into it.

\section{Discussion}\label{sec:discussion}

\subsection{Quantum-coherent readout}

All of the simulations in Sections \ref{sec:circuitsimulation}, \ref{sec:CCM},
and \ref{sec:mixedqucl}
describe deterministic fluxon evolution and thus result
in transmission probabilities of unity or zero, depending on the initial qubit state.
These deterministic dynamics are consistent with the readout
as a single-shot strong-projective measurement
--- in one qubit state the fluxon is transmitted with probability 1,
and in the other state, it is reflected.

To describe a probabilistic transmission regime,
where transmission amplitudes are between 0 and 1,
one needs to account for the quantum nature of the fluxon.
Although we have modeled the fluxons as purely classical,
as is justified by their large mass and as is done in other studies too \cite{FedorovETAL2007, Ustinov2013, Ustinov2014},
we can make some general observations about their quantum behavior,
based on the analysis of Ref.~\onlinecite{AveRabSem2006}.
The authors formulate several constraints for quantum-coherent readout,
and we evaluate these constraints for the LJJ parameters used in our study.

The initial probability distribution of a fluxon's center position $X$
may be characterized by a wave function $\Psi(X) e^{\ui k X}$,
where the wave vector $k = 8 E_0 v/(\hbar c^2) = 8 v/(\beta^2 \lambda_J c)$
is determined by the fluxon speed $\dot X = v$.
From the particle-like nature of each fluxon
follow two conditions on the wave function:
(i)
If $\Psi(X)$ has a width $\sigma_X = \xi$,
then the width of the distribution $\Psi(k)$ in momentum space is
$\sigma_k \sim 1/\xi$.
The particle-like nature of the fluxon implies that $\sigma_k$
needs to be small compared with $k_0$, resulting in the condition
$\xi/a \gg \beta^2 c \lambda_J/(8 v a)$.
For the values in our simulations, $\beta^2 = 0.4$, $v/c=0.6$
and $a/\lambda_J=1/\sqrt{7}$,
this amounts to $\xi/a \gg 0.22$.
Note that in the time-delay readout, a more restrictive constraint is
$\sqrt{l/k_0} \ll \xi \ll l$ where $l$ is the length of the LJJ.
The lower limit results in the condition
$\xi/a \gg (\beta^2 l \lambda_J/(8 a^2))^{1/2}$,
which amounts to e.g. $\xi/a \gg 2.6 $ if $l = 50 a$.
(ii)
 As mentioned above, a classical condition for suppressed discreteness effects is
 $\lambda_J > a$, and a similar condition may be imposed on the de-Broglie wavelength
 of the fluxon $\lambda_{\text{dB}} = 2\pi/k > a$.
 In our simulation, $\beta^2=0.4$ and $v/c=0.6$, and
 the de-Broglie wave length is $\lambda_{\text{dB}} = 0.52 \lambda_J = 1.4 a$,
 which thus fulfills the criterion.
 On the other hand, one may recast the condition of negligible discreteness effects as
 an upper limit on the allowed fluxon velocity,
 $v/c < 2\pi \beta^2 \lambda_J/(8a)$.
 For our simulation parameters this upper limit is $v/c < 0.8$.

 Similar to our ballistic gates with classically resonant dynamics, we have used the relatively high initial velocity $v/c=0.6$,
 where the fluxon carries the kinetic energy
 $2 E_0 = 2\hbar\omega_J/\beta^2 = 5 \hbar\omega_J$.
 This should allow a fast readout that occurs in a time much shorter than $1 \uns$.
 However, it is more relativistic than considered in Ref.~\onlinecite{AveRabSem2006}.
 In our case, the kinetic energy lies above the minimum energy $\hbar \omega_J$
 of plasma waves in the LJJ,
 whereas Ref.~\onlinecite{AveRabSem2006} suggests kinetic energies below $\hbar \omega_J$,
 in order to suppress plasma wave generation.
 To suppress this loss mechanism,
 the fluxon speed would need to be below
 $v/c < \sqrt{1 - (1 + \beta^2/8 )^{-2}} = 0.3$ at $\beta^2=0.4$.
 This is certainly a valid point.
However, in our system the fluxon mainly loses energy by exciting oscillations in the interface JJs.
If $v/c < 0.3$ this energy would indeed be too small to radiate away from the interface in form of plasma waves.
Nevertheless, this loss mechanism in principle renders the fluxon scattering inelastic.
In our presented case, however, the energy converted into the local oscillation (the bounce)
is mostly restored as kinetic energy of the outgoing fluxon.
The resulting scattering is thus near-elastic,
where the fluxon retains over $93\%$ ($96\%$) of its initial energy for $n=0$ ($n=1$).

While the consequence of the bath for pulsed c-QED readout is understood \cite{WallraffPulseReadout}, a future study will be required to explore the role of damping on and after our readout.
Bath engineering studies of other QIS systems have explored state preparation, state stabilization, quantum steering, and measurement-induced dynamics \cite{Murch1, Murch2}.

\subsection{Transmission vs. time-delay readout}

In the readout of this study, information about the qubit state is contained in the transmission probability.
This {\em transmission mode} readout lends itself to
strong projective measurements, relying on strong qubit-fluxon coupling for state-dependent transmission or reflection.
Another readout mode was proposed in Ref.~\cite{AveRabSem2006} as {\em time-delay mode},
where information about the qubit state is contained in the qubit-state dependent
passage time of the transmitted fluxon.
This latter method seems more suited for weak measurement schemes,
using a comparably weak qubit-fluxon coupling.
Following the original proposal,
several studies have further analyzed \cite{AnnaHerr2007b, Ustinov2013, Kuzmin2015}
and experimentally tested \cite{Ustinov2014} the time-delay readout mode.

Naturally, the delay-time readout mode is strongly affected by jitter,
i.e. the (qubit-independent) variation of transmission times caused by the effect
of random noise on the fluxon motion.
Jitter thus imposes limits on readout fidelity \cite{Kuzmin2015}.
In contrast, the transmission mode readout is practically unaffected by jitter
and this is advantageous considering that
jitter can be a dominant error source in LJJ transmission \cite{PanGorKuz2012, HanLJJ1}.

{\mycolor
The qubit measurement occurs during the interaction between the fluxon and the qubit, mediated by a finite $\phiBB$-excitation.  
In Fig.~\ref{fig2}, the readout dynamics is shown as a function of time normalized by the Josephson frequency. 
For the slower process, the interaction time is $4$ Josephson periods, as seen  in panel (a.ii). 
Choosing the fabrication parameters of Table \ref{tab:params}(A), we find the interaction or quantum measurement time to be $T_M = 0.33 \pm 0.03 \uns$. This time can be compared to the integration time of cavity-based qubit readout, which occurs over a time of 100s $\uns$ for an optimized readout \cite{GoogleReadout}.

The proposed readout technique requires fluxon measurement with low error after the fluxon-based readout of the qubit. The fluxon detector likely must be placed at least $(v/c)\times\omega_J t \simeq 3$ Josephson penetration depths from the qubit, which causes a delay time of $t \simeq 0.07 \uns$. Furthermore, previous studies \cite{YorozuETAL2002, YorozuETAL2007} at $4 \,\text{K}$ found that a flip-flop cell has an average delay of $10 \ups$ and can have negligible error during the flip process. Although the Josephson period ($2 \pi/\omega_J$) of our proposed JJ fabrication in Table \ref{tab:params} is much larger than a comparable flip-flop, we estimate the total time of qubit readout and fluxon measurement at less than $1 \uns$. 
}

\subsection{Alternative readout dynamics}

The readout dynamics presented in Secs.~\ref{sec:classical}
and \ref{sec:mixedqucl} involves multi-bounce scattering when the qubit is in state
$n=0$.
This is best seen in the CC model of Fig.~\ref{fig2}(a),
where the trajectory undergoes multiple bounces off the CC potential
and ultimately evolves into the channel with $X_R \to -\infty$.
One advantage of this scattering is that it can be highly
energy-conserving, minimizing fluctuations at the interface.
However, qubit-state distinction can also be achieved with the interface structure
without multi-bounce scattering types.
Figures~\ref{fig4} in App.~\ref{app:alternativereadout}
show an example, where the fluxon is reflected for $n=0$
and simply transmitted for $n=1$. To achieve these different dynamics,
parameters have been set such that the two values of $\langle \phi_q \rangle_n$
generate different boundary conditions in the CC potentials
$U(X_L, X_R, \phi_q = \langle \phi_q \rangle_n)$.

{\mycolor
If the parameters of the qubit and interface are chosen differently, an incoming fluxon can exhibit unintended dynamics. 
For example, if the parameters are changed such that the incoming fluxon is reflected at the end of the left LJJ as an antifluxon, a $4\pi$-phase change will result at the interface (in the termination JJ of the left LJJ and in the $\phiBB$-interface JJ ($\phiBB=\phiL - \phiR$)). 
Such a large phase change can cause non-deterministic Landau-Zener transitions in the qubit, similar to a large and fast flux bias change. 
}

\section{Conclusion}

In this theoretical study with simulations, we find a fast single-shot readout of a qubit that appears suitable for an experimental demonstration.
The readout uses a single ballistic fluxon, which scatters at an interface
between two LJJs, where the qubit resides. The qubit is strongly coupled to the interface, and its state can determine the fluxon scattering result.
Here we concentrate on transmission-mode readout, where one qubit eigenstate
leads to transmission and the other leads to reflection. This appears to allow for rapid readout without an input microwave tone. 

For a standard fluxonium qubit biased near an integer multiple of $\Phi_0$, the fluxon undergoes a single bounce, yielding reflection for one qubit state and a couple of bounces yielding transmission for the other qubit state.
These dynamics are initially studied semiclassically in circuit simulations,
with a separate degree of freedom for all JJs of the circuit.
However, a collective-coordinate model reduces the dynamics to three degrees of freedom: one for the qubit and one for each of the two LJJs.
We perform quantum simulations of this model within a Born-Oppenheimer approximation,
using the large mass imbalance between the light qubit coordinate and the heavy LJJ coordinates.

Our study shows a much more favorable outlook for transmission-mode readout of a qubit than previous work. One apparent advantage is a specified realistic circuit that yields a sensitive readout in the transmission mode using only a single fluxon.
The speed of the readout is dependent on the frequencies of the involved JJs and, since it is comparable to the Josephson period in a qubit JJ,
it will naturally be faster than $1 \uns$.
The calculated backaction is only $\approx 0.1\%$ on a standard fluxonium qubit
and a lower amount on a heavy fluxonium. Since this readout is fast, it should also be explored for high-temperature and high-frequency qubits.

\section{Acknowledgements}

WW was supported by {\it Quantum Technologies Aotearoa},
a research program funded by the New Zealand Ministry of Business,
Innovation and Employment, contract number UOO2347.
KDO thanks D. Averin for related scientific discussions.

\begin{appendix}

\section{Steady states of the coupled LJJ-qubit system}
\label{app:system}

Here we analyze the classical steady states of the circuit
in the absence of a moving fluxon.
The analysis is similar to that for a classical gate with two LJJs and one storage loop,
found in Sec.~IIB of Ref.~\onlinecite{WusOsb2023_BSR}.

In general, through its strong coupling with $\phiBB$ and $\phiL,\phiR$,
a finite amplitude of the qubit phase $\phi_q$ gives rise to finite amplitudes
of $\phiL$ and $\phiR$.
These in turn create evanescent fields in the LJJs, localized at the interface.
Assuming small amplitudes $|\phi_{L,R}| \ll \pi$,
we parametrize these bound states in the left and right LJJ as
\begin{equation}\label{eq:ansatz_boundstate}
 \phi_k^{(l,r)} = \phi_{L,R} e^{-\mu a k} \qquad (k = 0,1,2,\ldots)
 \,,
\end{equation}
where $\mu$ is the inverse decay length,
$k$ denotes the JJ index away from the interface,
and
$\phiL := \phi^{(l)}_{k=0}$ and $\phiR := \phi^{(r)}_{k=0}$
are shorthand notations for the phases of the interface's termination JJs.
(In general, the left and right LJJ may be in different `vacuum states',
i.e. the phases of all JJs of one LJJ may be shifted by an integer multiple of $2\pi$
relative to the other.
Such a phase difference between left and right LJJ may arise during a polarity-inverting
scattering or when a fluxon gets trapped at the interface,
which occurs in our classical ballistic gate studies,
but not in the simulations of this work.)

Inserting the bound-state ansatz \eqref{eq:ansatz_boundstate}
into the Lagrangian, \Eq{eq:Lagrangian},
the system Hamiltonian becomes, in the limits of small amplitudes, $\phi_{L,R} \ll \pi$,
\begin{align}\label{eq:H3_boundstate}
 H_{\text{bs}}^{(3)}
 &= H_q + H_{\text{LJJ}}^{(2)} + \Vint \\
 \frac{H_{\text{LJJ}}^{(2)}}{E_0} &= \frac{\pi_{LR}^2 \omega_J^2}{2\Mlr E_0^2}
 + \frac{\pi_{B}^2 \omega_J^2}{2\Mb E_0^2}
 + \frac{L\lambda_J}{\Leff a} \phiLR^2 \nonumber \\
 &
 + \frac{L\lambda_J}{2 a} \left(\frac{1}{2\Leff} + \frac{1}{L_q}\right) (\phiBB)^2
 + \frac{\IJbb a}{I_c \lambda_J} [ 1 - \cos\phiBB ] \nonumber \\
 &
 + \frac{2(\IJab + \IJeff)a}{I_c \lambda_J} [ 1 - \cos\phiLR \cos(\phiBB/2) ]
 \nonumber
\end{align}
with the Hamiltonian $H_q(\phiq)$ of the isolated fluxonium
and the interaction $\Vint$ given in \Eq{eq:Hq_phiq} and \Eq{eq:Vint},
respectively.
We have labeled the number of degrees of freedom with the superscripts $(n)$, $n=2,3$.
Here, instead of the bound state amplitudes $\phiL, \phiR$,
we use the symmetrized amplitude $\phiLR := (\phiL + \phiR)/2$,
while the anti-symmetrized amplitude is given by $\phiBB = \phiL-\phiR$.
Their conjugate charge variables are
$\pi_{LR} = \Mlr E_0 \omega_J^{-2} \dot{\phi}_{LR}$,
$\pi_{B} = \Mb E_0 \omega_J^{-2} \dot{\phi}^{B}$,
and
$\pq = \Mq E_0 \omega_J^{-2} \dot{\phi}_{q}$,
with the dimensionless masses
$\Mlr = \frac{2(\CJab + \CJeff)}{C_J\lambda_J/a}$,
$\Mb  = \frac{\CJbb + (\CJab + \CJeff)/2}{C_J\lambda_J/a}$,
and
$\Mq  = \frac{\CJq}{C_J\lambda_J/a}$.
Each LJJ contribution is reduced to an effective JJ and an effective inductance,
both in parallel with the corresponding termination JJ ($\IJab, \CJab$),
and their $\mu$-dependent parameters are
\begin{eqnarray}
 \CJeff &:=& C_J / (e^{2\mu a} - 1) \\
 \IJeff &:=& I_c / (e^{2\mu a} - 1) \\
 \Leff &:=& L (e^{\mu a} + 1) / (e^{\mu a} - 1)
 \,.
\end{eqnarray}
In these expressions the inverse decay length $\mu$ of the bound state
is not yet determined. However, we can estimate $\mu$ from the condition that
the bound state fulfills the dispersion relation in the LJJ bulk,
$\omega^2 = \omega_J^2 + 2c^2/a^2 \left( 1 - \cosh(a\mu) \right)$.
Being interested in steady states of the interface,
we can set $\omega=0$ and obtain the estimate
$\mu = a^{-1} \cosh^{-1}\left( 1 + a^2/(2\lambda_J^2) \right)$.


We concentrate on the parameter regime where the interface degrees of freedom,
$\phiLR$ and $\phiBB$, are heavy and strongly confined by the potential,
as opposed to a lighter fluxonium mass and a shallow potential in $\phiq$-direction.
For example, in the parameter case of Fig.~\ref{fig1},
$(\Mlr, \Mb, \Mq) = (2.4, 4.8, 0.28)$,
and at the steady state $(\phiLR,\phiBB,\phiq) \approx (0,0,0)$,
the local frequencies of the potential are
$(\omega_{LR}, \omega_{B}, \omega_q) = \omega_J \cdot (1.1, 0.78, 1.2)$,
leading to the values
$\langle (\Delta \phiLR)^2 \rangle = (0.09 \pi)^2$,
$\langle (\Delta \phiBB)^2 \rangle = (0.07 \pi)^2$,
and $\langle (\Delta \phiq)^2 \rangle = (0.25 \pi)^2$
of the ground state uncertainties
$\langle (\Delta \phi_i)^2 \rangle = \beta^2 \omega_J/(2 M_i \omega_i)$.
Due to the strong confinement of $\phiLR$ and $\phiBB$,
the first eigenstates of $H_{\text{bs}}^{(3)} = H_q + H_{\text{LJJ}}^{(2)} + \Vint$
inherit the qualitative characteristics of the isolated qubit $H_q$,
where the ground state and first excited state lie in different potential wells.
This is illustrated in Fig.~\ref{fig1}(c),
showing the first eigenstates of 
\begin{equation}\label{eq:H2_boundstate}
H_{\text{bs}}^{(2)} = H_q + H_{\text{LJJ}}^{(1)} + \Vint
\,,
\end{equation}
in which we have reduced $H_{\text{LJJ}}^{(2)}(\phiLR,\phiBB) \to H_{\text{LJJ}}^{(1)}(\phiBB)$
by neglecting the contribution of $\phiLR$ and $\pi_{LR}$ in \Eq{eq:H3_boundstate}.

\section{Collective-coordinate model}\label{app:CCM}

Here we sketch the derivation of a collective coordinate model
for the circuit of Fig.~\ref{fig1}(a), leading to the results
discussed in Sec.~\ref{sec:CCM}.
The procedure is similar to the collective coordinate analysis for
the ballistic flip-flop and other RFL gates \cite{WusOsb2023_BSR,
WusOsb2020_RFL}.

\begin{figure*}[t]
\includegraphics[width=\textwidth]{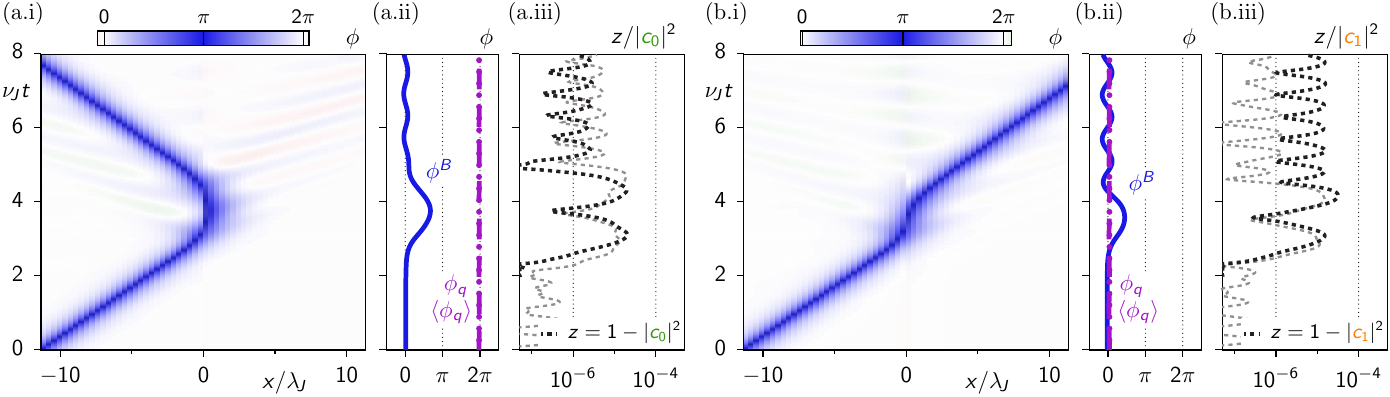}\\
\includegraphics[width=\textwidth]{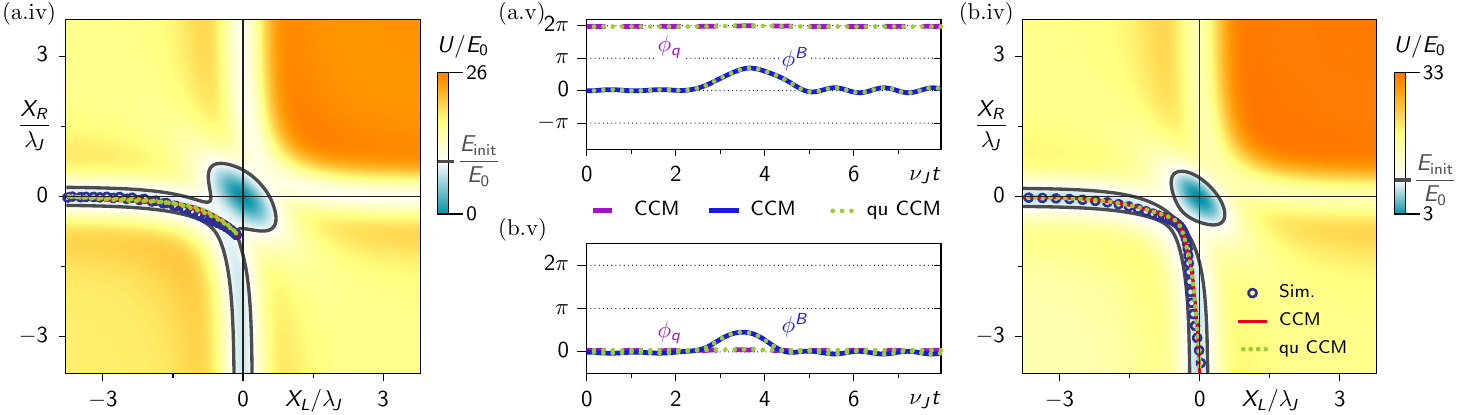}
 \caption{
  Simulated scattering results, including: LJJ phase evolution, trajectory in CC potentials, and dynamics of central gate junction $\phiBB$.
  The data structure is equivalent to those of Figs.~\ref{fig2} and ~\ref{fig3}.
 The fluxonium parameters are changed to $\phiext=1.2\pi$, $\IJq/I_c = 6.0$,
 $\CJq/C_J = 0.6$, $\Lq/L = 40$, and $\beta^2 = (4e^2/\hbar) \sqrt{L/C_J} = 0.4$,
 giving rise to the qubit transition energy of $\hbar \omega_{01} = 2.9 E_c^q = 0.25 E_0$.
 The interface parameters are changed to
 $C_J^B/C_J = 11.5$, $I_c^B/I_c = 6.7$, $\hat{C}_J/C_J = 0.75$,
 and $\hat{I}_c/I_c = 2.0$.
 }
 \label{fig4}
\end{figure*}

The starting point is the circuit Lagrangian,
\Eq{eq:Lagrangian} with the interface contribution, \Eq{eq:Lagr_intf}.
Inserting the mirror fluxon ansatz, \Eq{eq:fluxoncombination},
the LJJ contributions become
\begin{eqnarray}\label{eqA:L_CCM_LJJonly}
&& \frac{1}{E_0} \left(\mathcal{L}_l + \mathcal{L}_r \right)
= \sum_{i=L,R} \frac{m_0(X_i)}{2} \frac{\dot{X}_i^2}{c^2} - U_0(X_L,X_R)
\,, \\
\label{eqA:U0_CCM_LJJonly}
&& U_0 = \sum_{i=L,R} \Biggl\{
 \frac{4\lambda_J}{W} \left( 1 - \frac{2 z_i}{\sinh(2 z_i)} \right) \Biggr.\\
&&\hspace*{1.5cm} \Biggl.+ \frac{2 W}{\lambda_J}
 \tanh(z_i) \sech^2(z_i) \left[ 2 z_i + \sinh(2 z_i) \right] \Biggr\}
 \,, \nonumber  \\
\label{eqA:m0_CCM_LJJonly}
&& m_0(X_i) = \frac{8\lambda_J}{W} \left(1 + \frac{2 z_i}{\sinh(2z_i)} \right)
\,,
\end{eqnarray}
where $z_i = X_i/W$ ($i=L,R$).
To obtain these expressions, we have replaced the
LJJ sums in $\mathcal{L}_{l,r}$ by integrals,
based on the small discreteness, $a/\lambda_J \ll 1$.
We have evaluated all integrals with boundaries $(-\infty,0)$ and $(0,\infty)$,
which corresponds to including the interface's termination JJs as part of the LJJ.
To correct for this, the corresponding energies have to be subtracted
in the interface Lagrangian $\mathcal{L}_I$, \Eq{eq:Lagr_intf},
such that $\CJab \to \CJab - C_J$ and $\IJab \to \IJab - I_c$.
After inserting the ansatz \eqref{eq:fluxoncombination} also into $\mathcal{L}_I$,
the full system Lagrangian becomes
\begin{align}\label{eqA:Lagr_CCM}
\frac{\mathcal{L}}{E_0} &=
  \frac{m_{\!L}\dot{X}_{\!L}^2}{2c^2} + \frac{m_{\!R}\dot{X}_{\!R}^2}{2c^2}
+ m_{\!LR} \frac{\dot{X}_{\!L} \dot{X}_{\!R}}{c^2}
+ \frac{M_q}{2 \omega_J^2} (\dot{\phi}_{q})^2 \\
&\quad- U(X_{\!L}, X_{\!R}, \phiq) \nonumber
    \,.
\end{align}
Here, the interface modifies
the dimensionless mass of \Eq{eqA:m0_CCM_LJJonly} and
also contributes a mass-coupling term,
\begin{eqnarray}
\label{eqA:mXi}
&&m_i(X_i) = m_0(X_i) + \frac{\CJab-C_J + \CJbb}{C_J \lambda_J/a} (g_I(X_i))^2
\,,\\
\label{eqA:mXLXR}
&&m_{\!LR}(X_L,X_R) = \frac{\CJbb}{C_J \lambda_J/a} g_I(X_L) g_I(X_R)
\,,
\end{eqnarray}
where the factor $g_I(X_i) = 4 \left(\lambda_J/W\right) \sech(X_i/W)$
describes the local influence of the interface.
The dimensionless CC potential of \Eq{eqA:U0_CCM_LJJonly} is also modified,
leading to
\begin{align}
\label{eqA:Ufl_CCM}
\Ufl &= U_0 + \frac{\IJab-I_c+\IJbb}{I_c\lambda_J/a} u_1
 + \frac{\IJbb a}{I_c\lambda_J} u_2
 + \frac{L \lambda_J}{L_q a} u_s
 \,,
\end{align}
with the interface contributions to the potential,
\begin{align}
\label{eqA:u1_CCM}
 u_1 &= \sum_{i=L,R} 8 \sech^2(z_i) \tanh^2(z_i)
 \,,  \\
\label{eqA:u2_CCM}
 u_2 &= - \prod_{i=L,R} \left[ 8 \sech^2(z_i) \tanh^2(z_i) \right]
 \, \\
 &+ \prod_{i=L,R} \left[4 \sech(z_i) \tanh(z_i) \left( 1 - 2 \sech^2(z_i) \right) \right]
 \,,\nonumber \\
\label{eqA:u3_CCM}
 u_s
 &= \frac{1}{2} \left( \sigma(\phiL-\phiR) + \sigma \phiext\right)^2
 \,.
\end{align}
Combined with the qubit and interaction potentials $\Uq$ and $\Vint$
[cf.~\Eqs{eq:Uq_phiq} and \eqref{eq:Vint}]
the total potential in \Eq{eqA:Lagr_CCM} reads
\begin{align}
\label{eqA:U_CCM}
& U = \Ufl(X_L,X_R) + \Uq(\phiq)/E_0 + \Vint/E_0 \\
& \frac{\Uq(\phiq)}{E_0}
= \frac{\IJq a}{I_J \lambda_J} \left[ 1 - \cos \phiq \right]
 + \frac{L \lambda_J}{2 L_q a} (\phiq - \phiext)^2 \\
\label{eqA:Vint_CCM}
& \frac{\Vint(X_L,X_R,\phiq)}{E_0}
= - \frac{L \lambda_J}{L_q a} [\phiL - \phiR] (\phiq - \phiext) \\
&\hspace*{2.5cm}= - \frac{L \lambda_J}{L_q a} \sigma \left[ \sigma (\phiL-\phiR) \right] (\phiq - \phiext)
\,. \nonumber
\end{align}

\section{A readout case with smaller backaction}\label{app:alternativereadout}

Here we briefly present simulation results for another qualitative case of the same device,
with parameters given in Fig.~\ref{fig4}.
Compared with the fluxonium discussed in Sec.~\ref{sec:parameters},
the fluxonium is here even deeper in the `heavy fluxonium' regime \cite{SchusterETAL2021},
$E_c^q \approx E_L^q \ll E_J^q$,
such that another doublet (formed by states $n=2,3$) occupies the lowest two
wells in addition to the doublet of states $n=0,1$.
The qubit transition energy (of the isolated fluxonium)
is $\hbar \omega_{01} \approx 2.9 E_c^q \approx 0.25 E_0$.
Assuming the specific LJJ characteristics of App.~\ref{sec:specificfab},
the transition frequency becomes $\omega_{01}/2\pi = 8.76 \uGHz$.

Here, the fluxon reflects at the interface if the qubit is initialized in $n=0$,
Fig.~\ref{fig4}(a.i),
but transmits with only a short delay if it is initialized in $n=1$,
Fig.~\ref{fig4}(b.i).
Again, these dynamics are somewhat more intuitive within the CC model,
as shown in panels (iv) of Fig.~\ref{fig4}.
In case of $n=0$ the CC particle directly bounces once off the CC potential
near the constriction to the center well
and thus it reverts its direction.
In contrast in the case $n=1$, the center well energetically inaccessible
for the incoming fluxon which simply proceeds towards the transmission channel.
In this example, the different qubit states $n=0,1$ show a much more
pronounced difference in the CC potential $U(X_L, X_R, \phiq)$
(and of its dynamically relevant part $U(X_L, X_R, \phiq) - U_q = \Ufl + \Vint$),
compared with the very subtle difference in Fig.~\ref{fig3}.
This is because $E_L^q$ is much larger here, giving rise to
a much larger change of $\Vint(X_{L,R} \approx 0) \approx -E_L^q 2\pi (\phiq - \phiext)$
between the two fluxonium states:
$\Vint(X_{L,R} \approx 0, \langle \phiq \rangle_{1}) - \Vint(X_{L,R} \approx 0, \langle \phiq \rangle_{0})
\approx E_L^q (2\pi)^2$.

\begin{table*}[tb]
\renewcommand\arraystretch{1.2}
\caption{
Characteristic LJJ parameters and resulting fluxonium characteristics
for the readout cases
discussed in (A) the main text and (B) Appendix \ref{app:alternativereadout}, respectively.
The values assume JJ fabrication with $j_c = 0.1 \umicroA/ \umicrom^2$
and $c_J = 40 \ufF/ \umicrom^2$.
The LJJ parameters are chosen to be consistent with the parameters used in the numerics,
$L_J/L = 7$ and $\beta^2 = 0.4$ in the LJJs, and the specific
fluxonium parameters $\IJq/I_c$ and $\CJq/C_J$ of (A,B).
Case (A) assumes finite shunt capacitors $\Csh$ for the JJs in the LJJ
and that the qubit JJ is fabricated with the same value of $j_c$.
The JJ area $A$ and relative shunt capacitance $\Csh/(c_J A)$
are then calculated from \Eqs{eq:JJarea_caseA} and \eqref{eq:Csh}, respectively.
Case (B) assumes $\Csh=0$ and $j_c^q/j_c > 1$ for the qubit JJ,
where $A$ follows from \Eq{eq:JJarea_caseB}.
In both cases, the large required inductances $L$ can be fabricated using high-kinetic inductance superconductors.
}
\begin{ruledtabular}
\begin{tabular}{c|ccccccc|cccccc}
& $A$& $I_c$ & $\Csh/C_J^{\text{bare}}$ & $C_J$ & $L$ & $\omega_J/2\pi$  & $E_0/h$
& $j_c^q/j_c$
& $\frac{E_J^q}{h}$ & $\frac{E_c^q}{h}$ &$\frac{E_L^q}{h}$ & $\omega_{01}/2\pi$ & $\omega_{01}^{\text{min}}/2\pi$ \\
& [$\umicrom^2$] & [$\umicroA$] & & [$\ufF$] & [$\unH$] & [$\uGHz$] & [$\uGHz$]
& & [$\uGHz$] & [$\uGHz$] & [$\uGHz$] & [$\uGHz$] & [$\uMHz$] \\
\hline
(A) & 0.23 & 0.023 & 0.32 & 12.1 & 2.05 & 12.1 & 30.1
& 1 & 11.2 & 2.2 & 0.34 & 5.16 & 116 \\
(B) & 0.26 & 0.026 & 0 & 10.6 & 1.78 & 13.9 & 34.7
& 9 & 78.6 & 3.1 & 2.3 & 8.76 & 0.36
\end{tabular}
\end{ruledtabular}
\label{tab:params}
\end{table*}

{\mycolor
Both the classical circuit simulation and the classical CC simulation are robust under even relatively large deviations $\Delta \phiq(t=0)| \leq 0.6 \pi$ of the initial fluxonium phase $\phiq(t=0)$ from $ \langle \phiq \rangle_n$. 
This significantly exceeds the quantum uncertainties $\sqrt{ \langle (\Delta \phiq)^2 \rangle } \approx 0.17 \pi$ of the first two fluxonium states. 
Again, this demonstrates that our classical simulations are consistent with expected qubit fluctuations. 
Moreover, it indicates robustness of the readout method against these fluctuations.
}

Since the external flux here exceeds half a flux quantum,
$\phiext = 1.2 \pi > \pi$,
states $n=0,1$ are localized such that
$\langle \phi_q \rangle_{n=1} - \langle \phi_q \rangle_{n=0} \approx -2\pi$.
This is in contrast to the fluxonium constellation discussed in the main text,
where $\langle \phi_q \rangle_{n=1} - \langle \phi_q \rangle_{n=0} \approx 2\pi$,
cf.~Fig.~\ref{fig2}.
We note that for a symmetry-related fluxonium potential, which is obtained by
setting $\phiext = 0.8\pi < \pi$ on the other side of the
symmetry point $\phiBB + \phiext = \pi$,
the readout dynamics would be fully equivalent to those shown in Fig.~\ref{fig4}
if instead of a fluxon an antifluxon were employed,
i.e. a fluxon with inverted polarity $\sigma$.
This follows from the symmetry of \Eq{eq:Lagrangian} under the transformation
$\phi_k^{(l,r)} \to -\phi_k^{(l,r)}$, $\phiBB \to -\phiBB$, $\phi_q \to 2\pi -\phi_q$,
and $\phiext \to 2\pi - \phiext$.
The incoming antifluxon would then drive $\phiBB$ towards a negative amplitude,
i.e. moving further away from the fluxonium symmetry point.

From both the backaction analysis of Sec.~\ref{sec:backaction}
and the mixed-quantum classical CC dynamics of Sec.~\ref{sec:mixedqucl}
we again calculate the expansion coefficients $c_m(t)$
in the instantaneous eigenstates of the driven fluxonium.
The resulting weight
$\sum_{m \neq n} |c_{m}(t)|^2$ of all states that were not initially excited
is shown in panels (iii) of Fig.~\ref{fig4}.
Here the asymptotic value at $t \to \infty$ is with
$\sum_{m \neq n} |c_{m}(t)|^2/|c_{n}(t)|^2 \lesssim 4 \cdot 10^{-5}$
much smaller than for the previous readout example,
and even during the scattering it remains below $10^{-4}$.
In particular, repopulation into the other qubit state is strongly suppressed
due to the small tunneling rate between the wells.
This case therefore shows lower backaction on the qubit than the typical fluxonium studied in the main text.


\section{LJJ and qubit characteristics for specific fabrication}
\label{sec:specificfab}

In the simulations of the qubit readout, all quantities are dimensionless,
being scaled by the characteristic LJJ units.
Thus, apart from the value of $a/\lambda_J$ and of $\beta^2 = 4e^2/\hbar \sqrt{L/C_J}$,
the LJJ characteristics $(I_c, C_J, L)$ themselves remain unspecified in the numerical calculations.
Similarly, the qubit and interface parameters are only specified relative to the LJJ
parameters, e.g.~$\IJq/I_c$, $\CJq/C_J$, and $L_q/L$.
Here we specify a particular fabrication
and evaluate the resulting LJJ and qubit characteristics.


Readout cases discussed in (A) the main text and (B) App.~\ref{app:alternativereadout} have $\omega_J^q/\omega_J = (\IJq C_J/(I_c \CJq))^{1/2} > 1$,
implying that either the JJs in the LJJ are capacitively shunted
or that the qubit JJ is fabricated with larger critical current density than the LJJ and interface JJs.

We assume the JJs are fabricated with a critical current density of $j_c = 0.1 \umicroA/ \umicrom^2$
and a capacitance per area of $c_J = 40 \ufF/ \umicrom^2$.
This $j_c$-value can be realized with annealed Nb trilayer JJs \cite{SchusterFab1}.
We want to find the LJJ characteristics ($I_c, C_J, L$), which
are consistent with our chosen values of $L_J/L=\lambda_J^2/a^2 = 7$
and of $\beta^2 = \frac{4e^2}{\hbar} \sqrt{\frac{L}{C_J}} = 0.4$.

Firstly, assuming that the qubit JJ is fabricated with the same $j_c$-value,
the JJs in the LJJ have to be shunted with capacitors of size $\Csh$.
The total cacapitance of each JJ is $C_J = C_J^{\text{bare}} + \Csh$,
where $C_J^{\text{bare}} = c_J A$ and $A$ is the JJ area.
Using that $I_c = j_c A$ and $\IJq/\CJq = j_c/c_J$,
the relative shunt strength, $\Csh/(c_J A)$, is determined by the specific
value of $\omega_J^q/\omega_J$:
\begin{equation}\label{eq:Csh}
  \frac{\Csh}{C_J^{\text{bare}}} = \frac{C_J - c_J A}{c_J A}
     =\frac{\IJq}{I_c} \frac{C_J}{\CJq} - 1
     \,,
\end{equation}
Inserting \Eq{eq:Csh} into the equation defining $\beta^2$ and rearranging, one finds
the consistent value of the JJ area:
\begin{equation}\label{eq:JJarea_caseA}
 A = \sqrt{ \frac{8e^3}{\hbar} \frac{1}{j_c c_J (L_J/L) \beta^4 (\IJq/I_c)(C_J/\CJq)} }
 \,.
\end{equation}
Using this formula we find, for the readout case discussed in the main text,
the LJJ and qubit characteristics listed in row (A) of Table \ref{tab:params}.
With the resulting value of the LJJ energy scale $E_0$ we can then evaluate
the values of the fluxonium characteristics $E_J^q, E_c^q, E_L^q$ and $\omega_{01}$ 
(at the chosen bias $\phiext = 0.2\pi$) and $\omega_{01}^{\text{min}}$
at the half-flux quantum bias $\phiext = \pi$.

For the readout case discussed in App.~\ref{app:alternativereadout}
the LJJ-parameters for this type of fabrication would be demanding, requiring a
very small JJ area $A = 0.084 \umicrom^2$ and large participation ratio $\Csh/C_J^{\text{bare}}=9$.
In this case it is thus favorable to set $\Csh=0$ while fabricating the
qubit JJ with larger critical current density $j_c^q > j_c$,
determined by the ratio $j_c^q/j_c = I_c^q C_J/(I_c C_J^q)$.
The JJ areas in the LJJ can again be determined from the definition of $\beta^2$,
\begin{equation}\label{eq:JJarea_caseB}
 A = \sqrt{ \frac{8e^3}{\hbar} \frac{1}{j_c c_J (L_J/L) \beta^4} }
 \,.
\end{equation}
Using this formula we find, for the readout case discussed in App.~\ref{app:alternativereadout},
the LJJ and qubit characteristics listed in row (B) of Table \ref{tab:params}.

\end{appendix}


\begin{thebibliography}{0}%
\makeatletter
\providecommand \@ifxundefined [1]{%
 \@ifx{#1\undefined}
}%
\providecommand \@ifnum [1]{%
 \ifnum #1\expandafter \@firstoftwo
 \else \expandafter \@secondoftwo
 \fi
}%
\providecommand \@ifx [1]{%
 \ifx #1\expandafter \@firstoftwo
 \else \expandafter \@secondoftwo
 \fi
}%
\providecommand \natexlab [1]{#1}%
\providecommand \enquote  [1]{``#1''}%
\providecommand \bibnamefont  [1]{#1}%
\providecommand \bibfnamefont [1]{#1}%
\providecommand \citenamefont [1]{#1}%
\providecommand \href@noop [0]{\@secondoftwo}%
\providecommand \href [0]{\begingroup \@sanitize@url \@href}%
\providecommand \@href[1]{\@@startlink{#1}\@@href}%
\providecommand \@@href[1]{\endgroup#1\@@endlink}%
\providecommand \@sanitize@url [0]{\catcode `\\12\catcode `\$12\catcode `\&12\catcode `\#12\catcode `\^12\catcode `\_12\catcode `\%12\relax}%
\providecommand \@@startlink[1]{}%
\providecommand \@@endlink[0]{}%
\providecommand \url  [0]{\begingroup\@sanitize@url \@url }%
\providecommand \@url [1]{\endgroup\@href {#1}{\urlprefix }}%
\providecommand \urlprefix  [0]{URL }%
\providecommand \Eprint [0]{\href }%
\providecommand \doibase [0]{http://dx.doi.org/}%
\providecommand \selectlanguage [0]{\@gobble}%
\providecommand \bibinfo  [0]{\@secondoftwo}%
\providecommand \bibfield  [0]{\@secondoftwo}%
\providecommand \translation [1]{[#1]}%
\providecommand \BibitemOpen [0]{}%
\providecommand \bibitemStop [0]{}%
\providecommand \bibitemNoStop [0]{.\EOS\space}%
\providecommand \EOS [0]{\spacefactor3000\relax}%
\providecommand \BibitemShut  [1]{\csname bibitem#1\endcsname}%
\let\auto@bib@innerbib\@empty
\end{thebibliography}%


\begin{thebibliography}{}


\bibitem{Switch2}
B.L.T.~Plourde, T.L.~Robertson, P.A.~Reichardt, T.~Hime, S.~Linzen, C.-E.~Wu, and J.~Clarke,
{\em Flux qubits and readout device with two independent flux lines},
Phys. Rev. B {\bf 72}, 060506 (2005).

\bibitem{Switch3}
M.H.~Devoret, A.~Wallraff, and J.M.~Martinis,
{\em Superconducting Qubits: A Short Review},
arXiv:cond-mat/0411174 (2004).

\bibitem{Cavity1}
A.~Wallraff, D.I.~Schuster, A.~Blais, L.~Frunzio, R.-S.~Huang, J.~Majer, S.~Kumar,
S.M.~Girvin, and R.J.~Schoelkopf,
{\em Strong coupling of a single photon to a superconducting qubit using circuit quantum electrodynamics},
Nature {\bf 431}, 162 (2004).

\bibitem{Cavity2}
I.~Chiorescu, Y.~Nakamura, C.J.P.M.~Harmans, and J.E.~Mooij,
{\em Coherent Quantum Dynamics of a Superconducting Flux Qubit}
Science {\bf 299}, 1869 (2003).

\bibitem{cQED}
A.~Blais, A.L.~Grimsmo, S.M.~Girvin, and A.~Wallraff,
{\em Circuit quantum electrodynamics},
Rev. Mod. Phys. {\bf 93}, 025005 (2021).

\bibitem{SchoelkopfEC}
M.~Reed, L.~DiCarlo, S.~Nigg, L.~Sun, L.~Frunzio, S.M.~Girvin, and R.J.~Schoelkopf,
{\em Realization of three-qubit quantum error correction with superconducting circuits},
Nature {\bf 482}, 382 (2012).

\bibitem{WallraffEC}
S.~Krinner, N.~Lacroix, A.~Remm, et al.,
{\em Realizing repeated quantum error correction in a distance-three surface code},
Nature {\bf 605}, 669 (2022).

\bibitem{GoogleEC}
Google Quantum AI and Collaborators,
{\em Quantum error correction below the surface code threshold},
Nature (2024).

\bibitem{ChalmersReadout}
L.~Chen, H.-X.~Li, Y.~Lu, et al.,
{\em Transmon qubit readout fidelity at the threshold for quantum error correction without a quantum-limited amplifier}
npj Quantum Inf {\bf 9}, 26 (2023).

\bibitem{msMemory}
M.~Reagor, W.~Pfaff, C.~Axline, et al.,
{\em Quantum memory with millisecond coherence in circuit QED},
Phys. Rev. B {\bf 94}, 014506 (2016).


\bibitem{AveRabSem2006}
D.V.~Averin, K.~Rabenstein, and V.K.~Semenov:
{\em Rapid ballistic readout for flux qubits},
PRB {\bf 73}, 094504 (2006).

\bibitem{Ustinov2013}
K.G.~Fedorov, A.V.~Shcherbakova, R.~Sch\"afer, and A.V.~Ustinov:
{\em Josephson vortex coupling to a flux qubit},
APL {\bf 102}, 132602 (2013).

\bibitem{Kuzmin2015}
I.I.~Soloviev, N.K.~Klenov, A.L.~Pankratov, L.S.~Revin, E.~Il'ichev, and L.S.~Kuzmin:
{\em Soliton scattering as a measurement tool for weak signals},
PRB {\bf 92}, 014516 (2015).

\bibitem{FedorovETAL2007}
A.~Fedorov, A.~Shnirman, G.~Sch\"on, and A.~Kidiyarova-Shevchenko:
{\em Reading out the state of a flux qubit by Josephson transmission line solitons},
PRB {\bf 75} 224504 (2007).

\bibitem{AnnaHerr2007b}
A.~Herr, A.~Fedorov, A.~Shnirman, E.~Il'ichev, and G.~Sch\"on:
{\em Design of a ballistic fluxon qubit readout}, Supercond. Sci. Technol. {\bf 20}, S450 (2007).

\bibitem{Ustinov2014}
K.G.~Fedorov, A.V.~Shcherbakova, M.J.~Wolf, D.~Beckman, and A.V.~Ustinov:
{\em Fluxon Readout of a Superconducting Qubit},
PRL {\bf 112}, 160502 (2014).

\bibitem{SETreadout}
S.~Serrano, M.~Feng, W.H.~Lim et al.,
{\em Improved Single-Shot Qubit Readout Using Twin rf-SET Charge Correlations},
PRX Quantum {\bf 5}, 010301 (2024).

\bibitem{JPM}
A.~Opremcak, C.H.~Liu, C.~Wilen, K.~Okubo, B.G.~Christensen, D.~Sank, T.C.~White, A.~Vainsencher, M.~Giustina, A.~Megrant, B.~Burkett, B.L.T.~Plourde, and R.~McDermott,
{\em High-Fidelity Measurement of a Superconducting Qubit Using an On-Chip Microwave Photon Counter},
Phys. Rev. X {\bf 11}, 011027 (2021).

\bibitem{MukhanovReadout}
L.~Di Palma, A.~Miano, P.~Mastrovito, and D.~Massarotti, M.~Arzeo, and G.P.~Pepe, F.~Tafuri, and O.~Mukhanov,
{\em Discriminating the Phase of a Coherent Tone with a Flux-Switchable Superconducting Circuit},
Phys. Rev. Lett. {\bf 19}, 064025 (2023).



 \bibitem{Manucharyan2009}
V.E.~Manucharyan, J.~Koch, L.I.~Glazman, M.H.~Devoret:
{\it Fluxonium: Single Cooper-Pair Circuit Free of Charge Offsets},
Science {\bf 326}, 113 (2009).

 \bibitem{Manucharyan2019}
L.B. Nguyen, Y.-H. Lin, A. Somoroff, R. Mencia, N. Grabon, and V.E. Manucharyan,
{\em High-coherence fluxonium qubit},
Phys. Rev. X {\bf 9}, 041041 (2019).

\bibitem{Mooij1999}
J.E.~Mooij et al.,
{\em Josephson persistent-current qubit},
Science {\bf 285}, 1036 (1999).

\bibitem{Orlando1999}
T.P.~Orlando et al.
{\em Superconducting persistent-current qubit},
Phys. Rev. B {\bf 60}, 15398 (1999).

\bibitem{Orlando2016}
F. Yan, S. Gustavsson, A. Kamal, J. Birenbaum, A.P. Sears, D. Hover,
T.J. Gudmundsen, D. Rosenberg, G. Samach, S. Weber, J.L. Yoder, T.P. Orlando,
J. Clarke, A.J. Kerman, and W.D. Oliver,
{\em The flux qubit revisited to enhance coherence and reproducibility},
Nat. Comm. {\bf 7}, 12964 (2016).

\bibitem{ZhangETAl2021}
H. Zhang, S. Chakram, T. Roy, N. Earnest, Y. Lu, Z.Huang, D.K. Weiss, J. Koch, and D.I. Schuster,
{\em Universal Fast-Flux Control of a Coherent, Low-Frequency Qubit},
Phys. Rev. X {\bf 11}, 011010 (2021).

\bibitem{ManucharyanETAL2021}
A. Somoroff, Q. Ficheux, R.A. Mencia, H. Xiong, R.V. Kuzmin, and V.E. Manucharyan,
{\em Millisecond coherence in a superconducting qubit},
arXiv:2103.08578 (2021).

\bibitem{Manucharyan2024}
R.A. Mencia, W.-J. Lin, H. Cho, M.G. Vavilov, V.E. Manucharyan,
{\em Integer fluxonium qubit},
PRX Quantum {\bf 5}, 040318 (2024).

\bibitem{SchusterETAL2021}
A. Gyenis, A. Di Paolo, J. Koch, A. Blais,  A.A. Houck, and D.I. Schuster,
{\em Moving beyond the Transmon: Noise-Protected Superconducting Quantum Circuits},
PRX Quantum {\bf 2}, 030101 (2021).

%




\bibitem{HanLJJ1}
H.~Cai, L.~Yu, W.~Wustmann, R.~Clarke, and K.D.~Osborn,
{\em Detection of low-energy fluxons from engineered long Josephson junctions for efficient computing},
https://arxiv.org/pdf/2406.15671 (2024).

\bibitem{WusOsb2023_BSR}
K.D.~Osborn and W.~Wustmann,
{\em Asynchronous Reversible Computing Unveiled Using Ballistic Shift Registers},
Phys. Rev. Applied {\bf 19}, 054034 (2023).

\bibitem{BaronePaterno}
A.~Barone and G.~Patern{\`o},
{\it Physics and Applications of the Josephson Effect},
Wiley, New York, (1982).

\bibitem{BaroneETAL1971}
A.~Barone, F.~Esposito, C.J.~Magee, and A.C.~Scott,
{\em Theory and applications of the sine-gordon equation},
La Rivista del Nuovo Cimento {\bf 1}, 227, (1971).

\bibitem{WusOsb2020_RFL}
W.~Wustmann and K.D.~Osborn,
{\em Reversible fluxon logic: Topological particles allow ballistic gates along one-dimensional paths},
Phys. Rev. B {\bf 101}, 0.14516 (2020).



\bibitem{BraKiv1998}
O.M.~Braun and Y.S.~Kivshar,
{\it Nonlinear dynamics of the Frenkel-Kontorova model},
Phys. Rep. {\bf 306}, 1 (1998).




\bibitem{Ustinov_Nature2003}
A.~Wallraff, A.~Lukashenko, J.~Lisenfeld, A.~Kemp, M.V.~Fistul, Y.~Koval,
and A.V.~Ustinov,
{\em Quantum dynamics of a single vortex},
Nature {\bf 425}, 155 (2003).


\bibitem{HKILJJ}
M.~Wildermuth, L.~Powalla, J.N.~Voss, Y.~Schön, A.~Schneider, M.V.~ Fistul, H.~Rotzinger, and A.V.~Ustinov,
{\em Fluxons in high-impedance long Josephson junctions},
Appl. Phys. Lett. {\bf 120}, 112601 (2022).

\bibitem{TiN1}
P.~Diener, H.G.~Leduc, S.J.C.~Yates, Y.J.Y.~Lankwarden, J.J.A.~Baselmans,
{\em Design and testing of Kinetic Inductance Detectors made of Titanium Nitride},
J. Low-Temp. Phys. {\bf 167}, 305 (2012).

\bibitem{TiN2}
H.M.I.~Jaim, J.A.~Aguilar, B.~Sarabi, Y.J.~Rosen, A.N.~Ramanayaka, E.H.~Lock, C.J.K.~Richardson, and K.D.~Osborn,
{\em Superconducting TiN Films Sputtered over a Large Range of Substrate DC Bias},
IEEE Trans. Appl. Supercond. {\bf 25}, 1 (2015).

\bibitem{TiN3}
C.~Joshi, W.~Chen,1 H.G.~LeDuc, P.K.~Day, and M.~Mirhosseini,
{\em Strong Kinetic-Inductance Kerr Nonlinearity with Titanium Nitride Nanowires},
Phys. Rev. Appl. {\bf 18}, 064088 (2022).

\bibitem{SEEQCfab}
D.~Yohannes, M.~Renzullo, J.~Vivalda, A.C.~Jacobs, M.~Yu, J.~Walter, A.F.~Kirichenko,
I.V.~Vernik, and O.A.~Mukhanov,
{\em High density fabrication process for single flux quantum circuits},
Appl. Phys. Lett. {\bf 122}, 212601 (2023).
https://doi.org/10.1063/5.0152552

\bibitem{Mergemon}
H.J.~Mamin, E.~Huang, S.~Carnevale, et al.,
{\em Merged-Element Transmons: Design and Qubit Performance},
Phys. Rev. Appl. {\bf 16}, 024023 (2021).

\bibitem{SchusterFab1}
A.~Anferov, K.-H.~Lee, F.~Zhao, J.~Simon, and D.I.~Schuster,
{\em Improved coherence in optically defined niobium trilayer-junction qubits},
Phys. Rev. Appl. {\bf 21}, 024047 (2024).

\bibitem{SchusterFab2}
A.~Anferov, S.P.~Harvey, F.~Wan, J.~Simon, D.I.~Schuster,
{\em Superconducting Qubits above 20 GHz Operating over 200 mK},
arXiv:2402.03031 (2024).


\bibitem{Coleman1975}
S. Coleman,
{\em Quantum sine-Gordon equation as the massive Thirring model},
Phys. Rev. D {\bf 11}, 2088 (1975).

\bibitem{SpectralWalls}
C. Adam, K.~Oles, T.~Romanczukiewicz, and A.~Wereszczynski,
{\em Spectral Walls in Soliton Collisions},
Phys. Rev. Lett. {\bf 122}, 241601 (2019).

\bibitem{Goldobin2002}
E. Goldobin, D. Koelle, and R. Kleiner,
{\em Semifluxons in long Josephson $0$-$\pi$-junctions},
Phys. Rev. B {\bf 66}, 100508(R) (2002).

\bibitem{DauxoisPeyrard}
T.~Dauxois and M.~Peyrard,
{\em Physics of Solitons}
(Cambridge University Press, Cambridge, 2006).

\bibitem{McLaughlinScott1978}
D.W.~McLaughlin and A.C.~Scott,
{\it Perturbation analysis of fluxon dynamics},
Phys. Rev. A {\bf 18}, 1652 (1978).

\bibitem{Rajaraman}
R.~Rajaraman,
{\em Solitons and Instantons}
(North-Holland, Amsterdam, 1989). 

\bibitem{Fadeev1978}
L.D.~Fadeev and V.E.~Korepin,
{\em Quantum theory of solitons},
Phys. Rep. {\bf 42}, 1 (1978).

\bibitem{commentBornOppenheimer}
The adiabatic approximation is related to the Born-Oppenheimer approximation
for calculating molecular spectra. In its first step, the kinetic energy of the
nuclei is neglected and the electronic energy calculated as the eigenstate
of the electronic Schr\"odinger equation, which parametrically depends on the
nuclear positions. By varying the nuclear positions one obtains a potential
energy surface which forms the potential for the nuclei whose Schr\"odinger equation
can then be solved to obtain total energy of the molecule.

\bibitem{WallraffPulseReadout}
R.~Bianchetti, S.~Filipp, M.~Baur, J.M.~Fink, M.~G\"oppl, P.J.~Leek, L.~Steffen, A.~Blais, and A.~Wallraff,
{\em Dynamics of dispersive single qubit read-out in circuit quantum electrodynamics},
Phys. Rev. A {\bf 80}, 043840 (2009).

\bibitem{Murch1}
P.M.~Harrington, E.J.~Mueller, and K.W.~Murch,
{\em Engineered dissipation for quantum information science},
Nat Rev Phys {\bf 4}, 660 (2022).

\bibitem{Murch2}
W.~Chen, M.~Abbasi, B.~Ha, S.~Erdamar, Y.N. Joglekar, and K.W.~Murch,
{\em Decoherence-Induced Exceptional Points in a Dissipative Superconducting Qubit},
Phys. Rev. Lett. {\bf 128}, 110402 (2022).


\bibitem{PanGorKuz2012}
A.L.~Pankratov, A.V.~Gordeeva, and L.S.~Kuzmin
{\em Drastic Suppression of Noise-Induced Errors in Underdamped Long Josephson Junctions},
Phys. Rev. Lett. {\bf 109}, 087003 (2012).

\bibitem{GoogleReadout}
A.~Bengtsson, A.~Opremcak, M.~Khezri, D.~Sank ,A.~Bourassa ,K.~J.~Satzinger, S.~Hong, C.~Erickson, B.~J.~Lester, K.~C.~Miao, A.~N.~Korotkov, J.~Kelly, Z.~Chen, and P.~V.~Klimov, 
{\em Model-Based Optimization of Superconducting Qubit Readout}, 
Phys. Rev. Lett. {\bf 132}, 100603 (2024).

\bibitem{YorozuETAL2002}
S.~Yorozu, Y.~Kameda, H.~Terai, A.~Fujimaki, T.~Yamada, and S.~Tahara,
{\em A single flux quantum standard logic cell library},
Physica C: Superconductivity {\bf 378--381}, 1471 (2002).

\bibitem{YorozuETAL2007}
S.~Iwasaki, M.~Tanaka, N.~Irie, A.~Fujimaki, N.~Yoshikawa, H.~Terai, and S.~Yorozu,
{\em Quantitative evaluation of delay time in the single-flux-quantum circuit},
Physica C: Superconductivity and its Applications {\bf 463--465}, 1068 (2007).


\end{thebibliography}
\end{document}